\documentclass[pre,twocolumn]{revtex4}
\usepackage{amssymb,amsmath}
\usepackage{graphicx}
\usepackage{subfigure}

\newcommand{\mb}[1]{\ensuremath{\mathbf{#1}}}
\newcommand{\h}[1]{\ensuremath{\hat{\mb{#1}}}}

\begin{document}
\title{Thermodynamic consistency of liquid-gas lattice Boltzmann simulations}
\author{A.J. Wagner} 
\affiliation{Department of Physics, North Dakota State University,
Fargo, ND 58105}

\begin{abstract}
Lattice Boltzmann simulations have been very successful in simulating
liquid-gas and other multi-phase fluid systems. However, the underlying
second order analysis of the equation of motion has long been known to
be insufficient to consistently derive the fourth order terms that are
necessary to represent an extended interface. These same terms are
also responsible for thermodynamic consistency, i.e. to obtain a
true equilibrium solution with both a constant chemical potential and
a constant pressure. In this article we present an equilibrium
analysis of non-ideal lattice Boltzmann methods of sufficient order to
identify those higher order terms that lead to a lack of thermodynamic
consistency. We then introduce a thermodynamically consistent
forcing method.
\end{abstract}

\maketitle

\section{Introduction}
The standard lattice Boltzmann approach leads to an ideal gas equation
of state. Several different approaches to simulate non-ideal fluids
with lattice Boltzmann have been introduced in the past. The main
application has been to simulate phase separation, although other
application like an increase in the speed of sound have also been
considered. There are two different philosophies to introduce the
non-ideal terms.  The first is guided by an atomistic picture and
local interactions are introduced through a forcing term
\cite{Rothman,chen2}. The second starts from the Navier-Stokes
equation for a non-ideal gas and tries to match the hydrodynamic limit
of the lattice Boltzmann equation with this macroscopic equation
\cite{Swift,chen}.

Whatever the underlying philosophy, each approach leads to some
non-ideal equation of state. Knowing the equation of state is
sufficient to predict the phase-behavior from equilibrium
thermodynamic arguments. The equilibrium densities are determined
through the Maxwell construction \cite{Callen}.  Many approaches fail
this test and the resulting phase-diagrams deviate from the
theoretical one.

For many simple applications where phase-separation is only required to
form a nearly immiscible system, these thermodynamic details may be of
limited importance. For simulations of phase-separation, however, such
details can be crucial. In this paper we will examine why lattice
Boltzmann approaches using a force to introduce the non-ideal equation
of state fail to obtain the correct thermodynamic behavior. We
explicitly identify the terms that lead to the non-thermodynamic
behavior of these methods.

The paper is organized as follows: we first identify the hydrodynamic
equations that we intend to simulate. We then discuss the equilibrium
behavior of these equations. We then introduce a general lattice
Boltzmann method which introduces non-ideal terms through either
forcing or pressure terms. The hydrodynamic limit of this approach is
presented. This allows us to identify how we can incorporate the
non-ideal pressure contributions by either incorporating a bulk force
or by altering the pressure moment of the equilibrium distribution.

These two different methods for non-ideal systems are equivalent to
the old Y. Chen method \cite{chen}, or the more recent extension by He
\textit{et al.} \cite{he1,he} when we use the bulk force and the
Holdych correction to the Swift model \cite{Holdych} when we directly
alter the pressure moment. However, we are not able to consistently
introduce surface tension effects at this point since those appear as
higher order derivatives that are not derivable by a second order
expansion \cite{Luo}.

These higher order derivatives, however, are necessary to achieve
thermodynamic consistency. We introduce a near equilibrium analysis of
sufficient order to consistently derive these higher order gradient
terms. This analysis uncovers correction terms for pressure and
forcing methods. Despite the fact that we only perform a 5th order
analysis we are able to identify the correction terms exactly when the
fluid is not advected with respect to the lattice. With this knowledge
we are then able to formulate a forcing method that achieves
thermodynamic consistency.


\section{Equations of motion for a non-ideal gas}
As a simple example for a non-ideal fluid we will examine a single
component fluid that can undergo a liquid-gas phase-transition. For
simplicity an isothermal system is considered.  Such a system
has an underlying free energy of the form
\begin{equation}
F=\int [f(\rho)+I(\nabla \rho,\nabla^2\rho, \ldots)]\; dx
\label{GenFree}
\end{equation}
where $\rho$ is the density, $f(\rho)$ is the bulk free energy and
$I(\nabla\rho,\nabla^2\rho,\ldots)$ is a gradient expansion of the interfacial free
energy. The lowest order term for the free energy is
$\frac{\kappa}{2}(\nabla\rho)^2$ and usually only this term is considered.  The
equations of motion for a non-ideal gas are given by the continuity
equation
\begin{equation}
\partial_t \rho+\nabla (\rho \h{u})=0,
\label{cont}
\end{equation}
$\h{u}$ is the velocity of the fluid, and
the Navier-Stokes equation of a non-ideal gas
\begin{equation}
\rho \partial_t \h{u}+ \rho \h{u}.\nabla \h{u} = \rho \nabla \mu +\nabla \sigma.
\label{NS}
\end{equation}
Here $\mu=\frac{\delta F}{\delta\rho}$ is the chemical potential of the non-ideal
gas and $\sigma=\nabla\h{u}+(\nabla\h{u})^T-\frac{2}{3}\nabla.\h{u}\mb{1}$ is the
Newtonian stress tensor.

Let us consider how this system will approach equilibrium. The
equilibrium will be time independent and the flow will be uniform,
\textit{i.e.} $\h{u}=$const. So the equation (\ref{NS}) reduces to 
\begin{equation}
\nabla\mu=0
\label{muconst}
\end{equation}
so that the chemical potential will be constant in equilibrium.

Condition (\ref{muconst}) only guarantees that the chemical potential
is constant in both phases. Bulk thermodynamics, however, requires
that the pressure is also constant in equilibrium. This is guaranteed
through the thermodynamic relation
\begin{equation}
\rho\nabla\mu = \nabla P
\label{Pcond}
\end{equation}
where $P$ is the pressure tensor. This pressure tensor is defined
through the properties that it is equal to the bulk pressure in the
absence of density gradients $P=p\mb{1}=[\rho\partial_\rho f(\rho)-f(\rho)]\mb{1}$ and
through condition (\ref{Pcond}) in the interfacial areas
\cite{thesis}. With this relation a uniform chemical potential is
equivalent to a divergence-free pressure tensor.

As an aside it is interesting to note that this statement is more
general than bulk thermodynamics. It can even be applied to finite
systems with curved interfaces. An example is an equilibrium drop that
will have a constant chemical potential, but the pressure inside the
drop will differ from the pressure outside by the Laplace pressure $\Delta
p=\frac{\sigma}{\rho}$, where $\sigma$ is the surface tension. Despite this
difference in the pressures inside and outside the drop the divergence
of the pressure tensor vanishes everywhere. For such a system with a
curved interface the equilibrium liquid and gas densities will differ
slightly from the bulk thermodynamic values \cite{drop}.

The connection between bulk thermodynamics and the
equilibrium condition $\rho\nabla\mu=\nabla P=0$ can be established by considering a flat
interface between the liquid and gas phases. Assume that this
interface is orthogonal to the x-axis. In this case $\nabla P=\partial_xP_{xx}
\mb{e}_x$ is just the derivative of a scalar function. Therefore it
follows from $\nabla P=0$ that in the bulk phase $p_l=p_g$. Therefore the
solution of the differential equation $\nabla P=\rho \nabla\mu=0$ for a flat
interface implies the standard bulk thermodynamic equilibrium conditions
$\mu_l=\mu_g$ and $p_l=p_g$.

Let us consider the condition $\rho \nabla \mu=0$ in some more detail. For
concreteness' sake let us assume that $I(\nabla\rho,\ldots)=\frac{\kappa}{2}(\nabla\rho)^2$. We
then obtain $\mu=\partial_\rho f(\rho)-\kappa \nabla^2\rho$ and $P=[\rho\partial_\rho f(\rho)-f(\rho)-\kappa
\rho\nabla^2\rho-(\kappa/2)(\nabla\rho)^2]\mb{1}+\kappa\nabla\rho\nabla\rho$. Then
the differential equation for a single flat interface becomes
\begin{equation}
\kappa \partial_x^3\rho = \partial_x \partial_\rho f(\rho)
\label{diff}
\end{equation}
subject to the boundary conditions that 
\begin{equation}
\lim_{x\to\pm \infty}\partial_x\rho=0.
\label{bc}
\end{equation}
Because (\ref{diff}) is equivalent to both $\nabla\mu=0$ and $\nabla P=0$ the
limiting values for the density will be the equilibrium densities
$\rho_g$ and $\rho_l$. This conclusion, however, is independent of the form
of the interfacial energy term $I(\nabla\rho,\nabla^2\rho,\ldots)$ and as long as the new
differential equation has a solution that fulfills the boundary
condition (\ref{bc}) the limiting densities $\rho_g$ and $\rho_l$ will be
the same.

The important corollary of this argument is that while there are many
possible forms of the chemical potential and corresponding pressure
tensor that lead to the correct bulk phase behavior, arbitrary 
derivative terms (that might arise because of unintentional higher
order corrections to the numerical method) will in general not be
derivable from an interfacial energy term $I(\nabla\rho,\ldots)$.  In such a situation
the differential equation (\ref{muconst}) can lead to bulk densities
that do not correspond to the equilibrium densities. These situations
are the main concern of this paper.

\section{Lattice Boltzmann}
The lattice Boltzmann method can be viewed as a discretization of the
Boltzmann equation. And in the same way that the Boltzmann equation
describes a gas that at long wavelength obeys the hydrodynamic
equations, the same is true for the lattice Boltzmann method. In the
lattice Boltzmann method both the space and velocity space is
discretized and the basic variables are the densities $f_i(x,t)$
associated with the velocity $v_i$. The hydrodynamic variables are
then the local density 
\begin{equation}
\rho(x,t)=\sum_i f_i(x,t)
\label{density}
\end{equation} 
and the momentum
\begin{equation}
\rho(x,t)\mb{u}(x,t)=\sum_if_i(x,t) \mb{v}_i.
\label{Udef}
\end{equation} 
Most lattice Boltzmann methods do not conserve energy and instead
enforce a constant temperature. One caveat is that \mb{u} is not
necessarily the local velocity, as we will see below.

The evolution equation for a non-ideal fluid can be
written as
\begin{eqnarray}
f_i(\mb{x}+\mb{v_i},t+1)&=&f_i(\mb{x},t)+F_i(\mb{x},t)+
\nonumber\\&&\frac{1}{\tau}(f_i^0(\rho)+A_i(\mb{x},t)-f_i(\mb{x},t)).
\label{LBeqn}
\end{eqnarray}
Here $f_i^0$ is the equilibrium distribution for the ideal gas. The
$A_i$ represent non-ideal contributions to the pressure tensor
\cite{Swift}, and $F_i$ are contributions of an external force. The
force can also be used to mimic interactions in a mean-field manner.

In lattice Boltzmann the Navier-Stokes equations are not discretized
directly; instead the moments of the equilibrium distribution as well
as $A_i$ and $F_i$ are chosen such that the momentum moment of a
second order expansion of the lattice Boltzmann equation will give the
desired Navier-Stokes equation (\ref{NS}). A dilemma occurs when the pressure
tensor itself contains second order derivative terms since these
terms are formally higher than second order in the Navier-Stokes
equations. These terms are not consistently derived in standard lattice
Boltzmann expansion techniques.

So the question arises: how can these terms be consistently
incorporated into a lattice Boltzmann method? One may consider simply
expanding the lattice Boltzmann equation to higher order, but this
will lead to Burnett level equations, which is not the desired
result. The reason that these higher order density derivatives appear
in the Navier Stokes equation is because the density derivatives are not
small near an interface and these terms are responsible for the
surface tension. 

It is difficult and rather cumbersome to extend the expansion of the
lattice Boltzmann equation to higher orders in a general way. So
instead we will consider an equilibrium (or at least stationary state)
interface instead and develop a higher order analysis for this
situation. Doing so uncovers the additional terms that lead to a lack
of thermodynamic consistency for forcing methods. Incorporating these
terms allows for thermodynamic consistency as will be discussed
below.

There are two expansion methods that are regularly used to derive the
hydrodynamic limit of the lattice Boltzmann equations: they are
referred to as the Taylor expansion and the Chapman-Enskog methods. Up
to second order the methods give identical results, but there is
debate about the equivalence of the two methods for higher order
results. In this paper we will utilize the first method.

To establish a starting point for a higher order expansion we will
first present the standard second order expansion. We will then show
how to choose the moments of $f_i^0$, $A_i$, and $F_i$ to
simulate a van der Waals gas.

\section{Second order expansion}
To obtain the hydrodynamic equations that govern the evolution of the
slow dynamics of the conserved quantities we use a Taylor expansion of
(\ref{LBeqn}) to second order. As usual we will only conserve mass,
defined through the density (\ref{density}) 
and momentum defined through (\ref{Udef}).

For this iso-thermal model, which does not
include a conserved energy moment, we require the knowledge of the
first three velocity moments of the equilibrium distribution
function. These are given by
\begin{eqnarray}
\sum_i f_i^0 &=& \rho \\
\sum_i f_i^0 (\mb{v}_i-\mb{u}) &=& 0\\
\sum_i f_i^0(\mb{v}_i-\mb{u})(\mb{v}_i-\mb{u})
&=& \rho\theta\mb{1}\\
\sum_i f_i^0(v_{i\alpha}-u_\alpha)(v_{i\beta}-u_\beta)
(v_{i\gamma}-u_\gamma)
&=& Q_{\alpha\beta\gamma}.
\end{eqnarray}
Galilean invariance would require the third order tensor $Q$ to
vanish. In most standard lattice Boltzmann methods this term does not
vanish, however, because these models contain too small a velocity
set. For these models $v_{ix}^3=v_{ix}$ which precludes the presence of the
third power of \mb{u} in the $\sum_i f_i^0v_{ix}^3$ moment. It also
fixes the temperature to be $\theta=\frac{1}{3}$. Therefore most models
have a $Q$ term given by the third order tensor
$Q=\rho\mathbf{u}\mathbf{u}\mathbf{u}$. The effects of the Galilean
invariance violation caused by this term become noticeable only at large
velocities $\mb{u}$ \cite{Li}.

The non-ideal contributions from the $A_i$ need to conserve mass and
momentum and the moments are given by
\begin{eqnarray}
\sum_i A_i^0 &=& 0 \\
\sum_i A_i^0 (\mb{v}_i-\mb{u}) &=& 0\\
\sum_i A_i^0(\mb{v}_{i\alpha}-\mb{u}_\beta)(\mb{v}_{i\beta}-\mb{u}_\beta)
&=& A_{\alpha\beta}\\
\sum_i A_i^0(v_{i\alpha}-u_\alpha)(v_{i\beta}-u_\beta)
(v_{i\gamma}-u_\gamma)
&=& A_{\alpha\beta}u_\gamma+A_{\alpha\gamma}u_\beta\nonumber\\&&+A_{\beta\gamma}u_\alpha. 
\end{eqnarray}
The forcing term $F_i$ needs to conserve mass, therefore
\begin{equation}
\sum_i F_i=0.
\end{equation}
It also needs to change the momentum by an amount $\mb{F}$,
therefore we choose
\begin{equation}
\sum_i F_i \left(\mb{v}_i-\mb{u}\right)=\mb{F}.
\end{equation}
The second moment of the forcing term is usually taken to be zero, but
we leave a general term $\Psi$ which we will use later to contain a
correction term.
\begin{equation}
\sum F_i \left(\mb{v}_i- \mb{u}\right) 
\left(\mb{v}_i- \mb{u}\right)
=\Psi.
\end{equation}
At this point it is worth while to note that the distinction between
the $A$ and the $\Psi$ terms is somewhat artificial. Both terms enter the
lattice Boltzmann equation in the same way, so that instead of $\Psi$ we
can insert the term $A=\tau\Psi$ and vice versa. We distinguish between
these terms to connect our analysis to established methods. Pressure
methods \cite{Swift,Holdych} only use $A$ and forcing methods
\cite{chen,he1,he,Luo,Guo} only $F$ and $\Psi$.

Now we need to derive the hydrodynamic limit of the lattice Boltzmann
equation (\ref{LBeqn}) to link this method to the hydrodynamic equations
(\ref{cont}), (\ref{NS}) which we want to simulate.

\subsection{The hydrodynamic limit}
While the Taylor expansion method is in principle well known we will
present it here again because the higher order analysis presented
later in this paper uses the results and techniques of this approach.
The main premise of the Taylor expansion approach is that the
distribution function can be expressed by a Taylor expansion
\begin{eqnarray}
f_i(x+v_i\Delta t,t+\Delta t)
&=&\sum_k \frac{(\Delta t)^k}{k!}D^kf_i(x,t).
\end{eqnarray}
where we have defined the derivative operator $D_i=(\partial_t+v_i\nabla)$. For
this expansion to be useful we must make the assumption that
derivatives are small. One could phrase the same argument in terms of
$\Delta t$, but it will be convenient to set $\Delta t=1$ in (\ref{LBeqn}). To
order our terms we will therefore use the derivatives as a small
parameter $\epsilon$ so that $O(\partial^n)=O(\epsilon^n)$. We obtain for
(\ref{LBeqn}) to second order 
\begin{equation}
D_i f_i + \frac{1}{2} D_i^2 f_i+F_i+ O(D^3)=\frac{1}{\tau}(\hat{f}_i^0-f_i),
\label{expand}
\end{equation}
where we have introduced $\hat{f}_i^0=F_i^0+A_i$.
This relates non-local $f_i$ to local $f_i$ and their
derivatives. However, the $f_i$ remain unknown and we only know the
$\hat{f}_i^0$ in terms of the macroscopic quantities. We can use
(\ref{expand}) to express the $f_i$ in terms of the equilibrium
distribution and higher order derivatives:
\begin{eqnarray}
f_i&=&\hat{f}_i^0-\tau F_i-\tau D_i f_i+ O(\partial^2)
\nonumber\\&=&\hat{f}_i^0-\tau F_i-\tau D_i(\hat{f}_i^0-\tau F_i)+ O(\partial^2).
\label{fzero_eqn}
\end{eqnarray}
Using this approximation we can now express the lattice Boltzmann
equation purely as a differential equation in terms of the equilibrium
distribution and the collision term: 
\begin{eqnarray}
F_i+D_i (\hat{f}_i^0 -\tau F_i)- \left(\tau-\frac{1}{2}\right)D_i^2
 (\hat{f}_i^0-\tau F_i)
\nonumber\\
=\frac{1}{\tau}(\hat{f}_i^0-f_i)+ O(\partial^3).
\label{expand_equil}
\end{eqnarray}
Taking the zeroth order velocity moment $\sum_i$(\ref{expand_equil}) we
obtain (borrowing the Euler level terms, \textit{i.e.} the terms of
$O(\partial)$, of the momentum equation (\ref{NSeqn})) the continuity equation:
\begin{equation}
\partial_t\rho+\nabla \left(\rho\mathbf{u}-\frac{1}{2} \mb{F}\right)=O(\partial^3).
\end{equation}
This leads us to identify the mean fluid velocity as
\begin{equation}
\hat{\mb{u}}=\mb{u} -\frac{1}{2\rho}\mb{F}.
\label{uhatdef}
\end{equation}
We then obtain the continuity equation
\begin{equation}
\partial_t \rho +\nabla (\rho \hat{\mb{u}})=O(\partial^3).
\end{equation}
Taking the first order velocity moment $\sum_iv_i$(\ref{expand_equil}) we
obtain the Navier-Stokes level equation:
\begin{equation}
\rho \partial_t\hat{\mb{u}}+\rho \hat{\mb{u}}.\nabla \hat{\mb{u}}=
-\nabla(\rho\theta+A)+F+\nabla\sigma+\nabla R  + O(\partial^3)
\label{NSeqn}
\end{equation}
where the Newtonian stress tensor $\sigma$ is given by
\begin{equation}
\sigma=\nu \rho (\nabla\h{u}+(\nabla\h{u})^T)
\label{sigma}
\end{equation}
and unphysical terms have been collected in the remainder tensor
\begin{equation}
R=\tau \Psi  -3\nu [\h{u}\nabla.A+(\h{u}\nabla.A)^T+\h{u}.\nabla A \mb{1}+\nabla Q]+ O(\partial^2).
\label{eqnR}
\end{equation}
The kinematic viscosity is given by $\nu =(\tau-\frac{1}{2})\theta$.  Note that
while we wrote \h{u} in equations (\ref{sigma}) and (\ref{eqnR}) it
cannot be distinguished from \mb{u} here because the forcing term is
$O(\partial)$ and therefore $\h{u}=\mb{u}+O(\partial)$. We also note that most of
the unphysical terms in (\ref{eqnR}) violate Galilean invariance
\cite{Holdych,Li}.

\subsection{Forcing and Pressure methods for the Non-ideal gas}
If we set both $A$ and $F$ to zero we obtain the Navier-Stokes
equation for an ideal gas.  To obtain the Navier Stokes equation for a
non-ideal gas previous research has identified two different
strategies. One option is to use the forcing term to introduce the
non-ideal contribution to the equation of state\cite{chen}. The form presented here was first presented
in the elegant work of He \textit{et al.} \cite{he1,he}.  In this case,
which we will refer to as the forcing method, we define
\begin{eqnarray}
F &=& -\nabla.P^{nid},\nonumber\\
\Psi &=& 0,
\label{forcing1}\\
A &=& 0,\nonumber
\end{eqnarray}
where $P^{nid}=P-n\theta \mb{1}$ is the non-ideal part of the pressure
tensor.  The second approach is based on the idea of including the
non-ideal pressure in the second moment of the equilibrium
distribution\cite{Swift,Holdych}. To do this we define
\begin{eqnarray}
F &=& 0,\nonumber\\
\Psi &=& 0,\label{pressure1}\\
A &=& P-\rho\theta \mb{1}+\nu (\h{u}\nabla\rho+(\h{u}\nabla\rho)^T+\h{u}.\nabla\rho \theta \mb{1}),
\nonumber
\end{eqnarray}
where the $\nu$ terms have been introduced by Holdych \textit{et al.}
\cite{Holdych} and later by Inamuro \textit{et al.} \cite{Inamuro}. To
do this we have to use the near-equilibrium approximation of $A=-\rho
\theta+O(\epsilon)$, as discussed by Kalarakis \cite{Kalarakis}.

\begin{figure}
\subfigure[]{
\includegraphics[width=0.8\columnwidth]{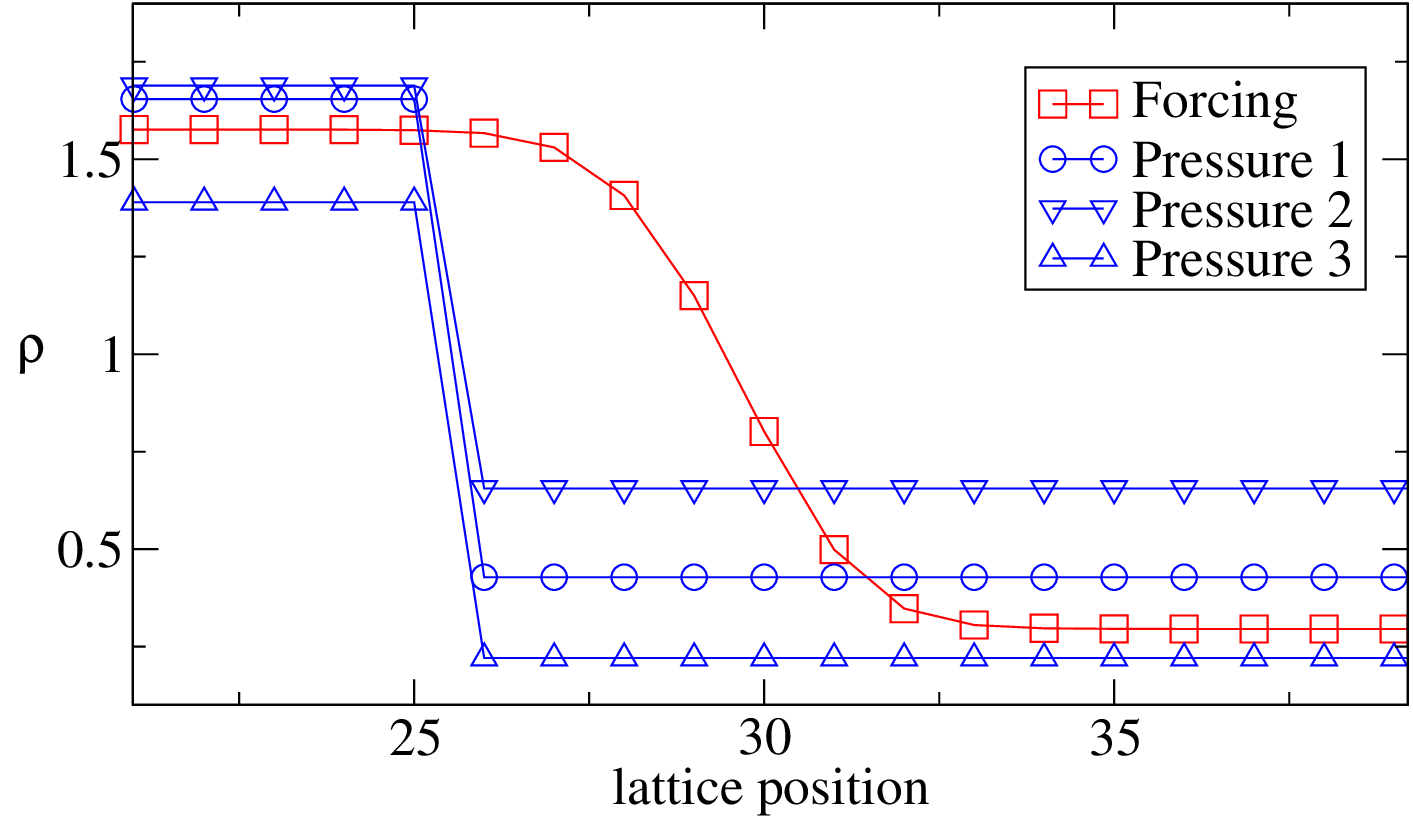}
}
\subfigure[]{
\includegraphics[width=0.8\columnwidth]{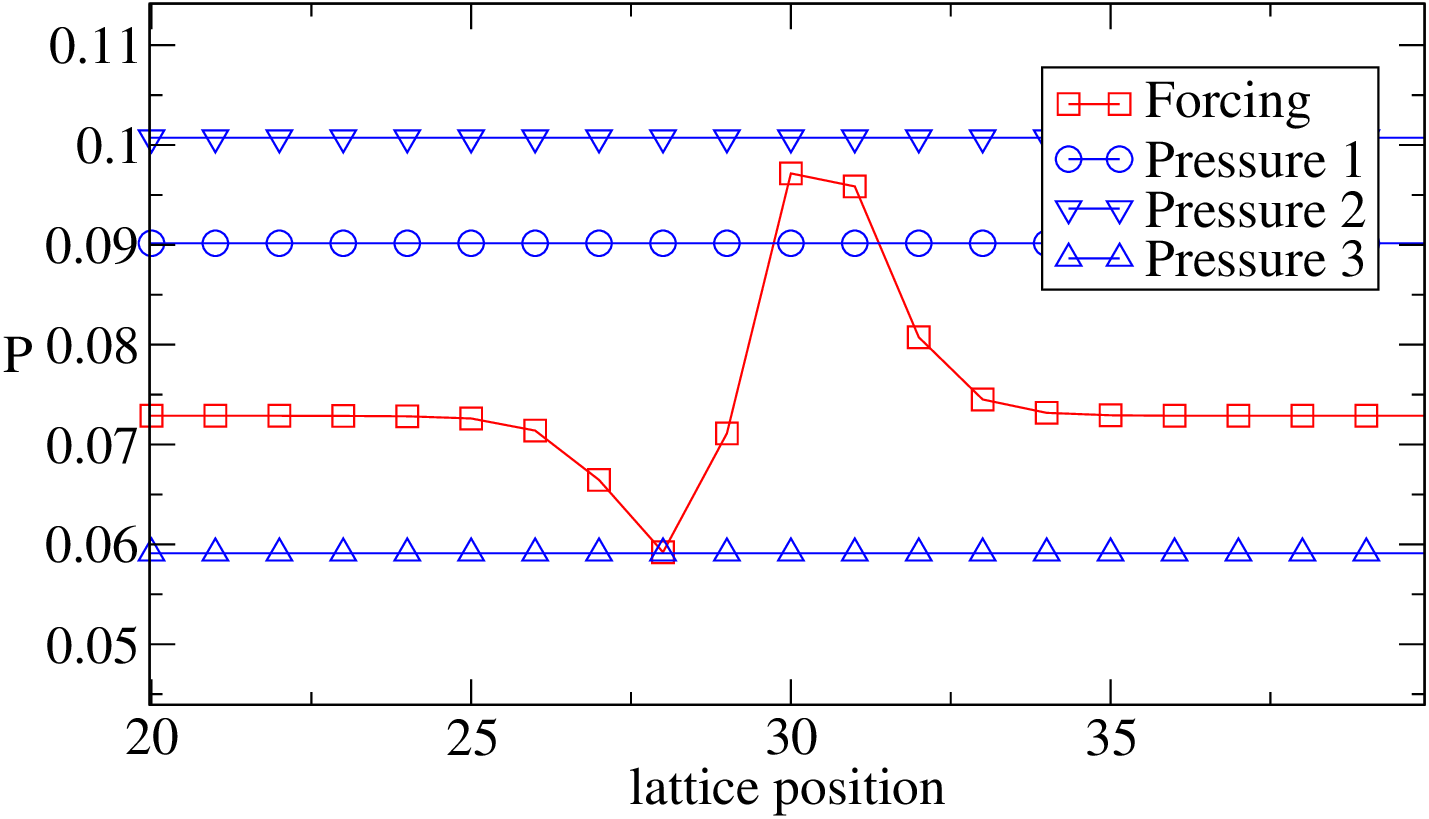}
}
\subfigure[]{
\includegraphics[width=0.8\columnwidth]{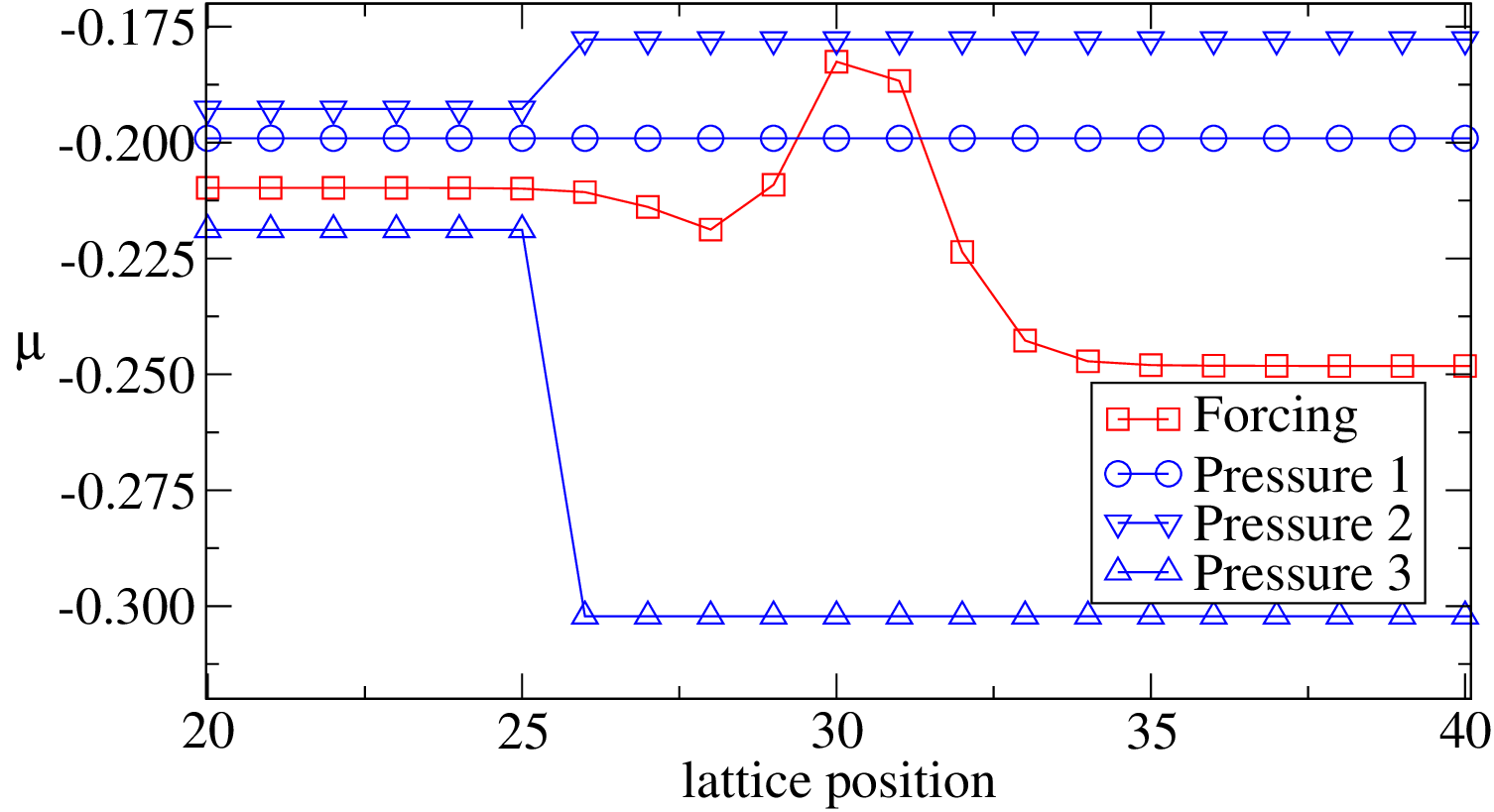}
}
\subfigure[]{
\includegraphics[width=0.8\columnwidth]{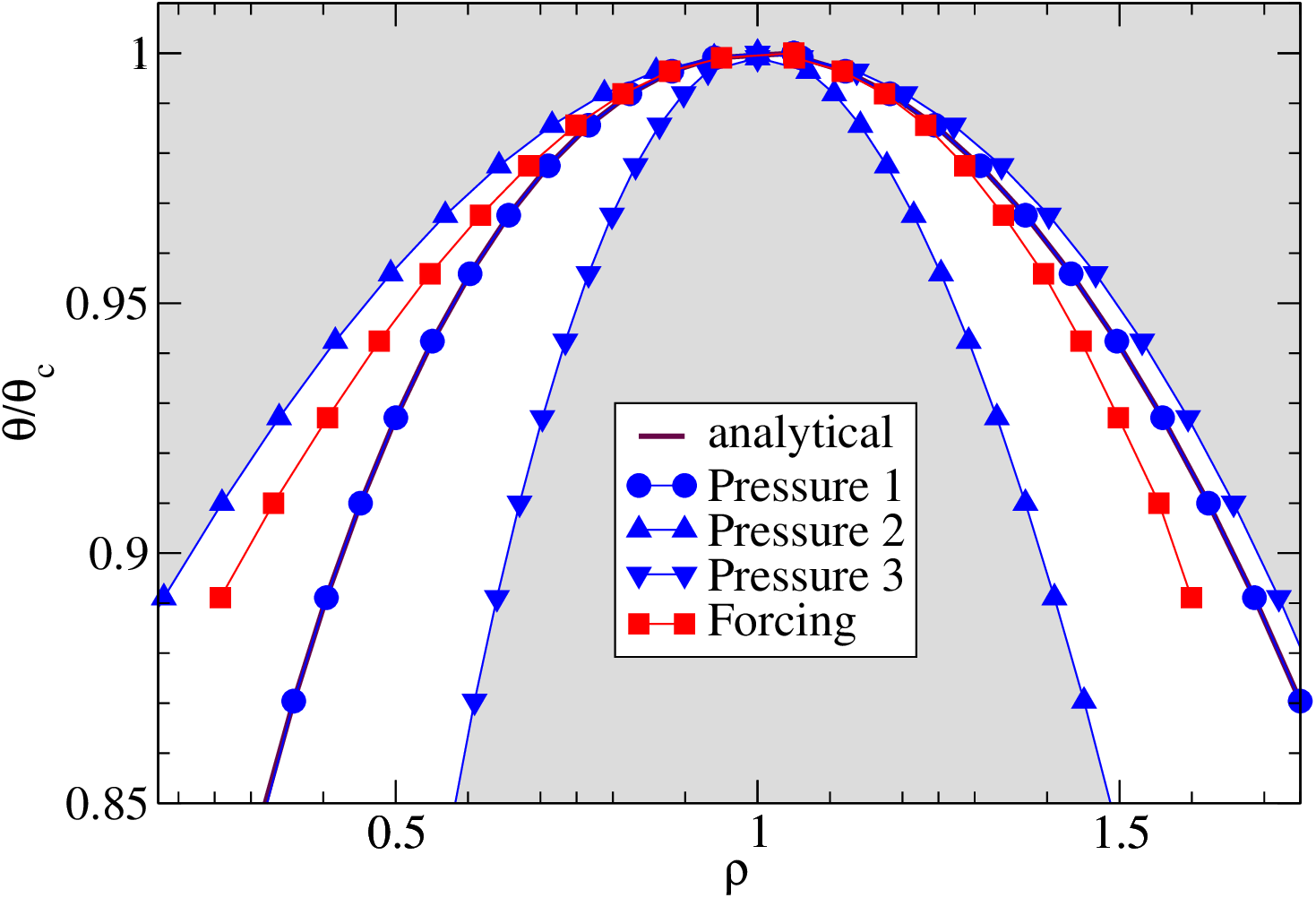}
}
\caption{Comparison of the profile (a) and the phase diagram (b) for
  the van der Waals pressure of equation (\ref{VdW}) obtained by the
  forcing method of equation (\ref{forcing1}) with the pressure method
  from equation (\ref{pressure1}). There is no unique solution to the
  pressure method. The unique solution of the forcing method is not in
  agreement with the theoretical phase diagram. The critical temperature
  used in (a-c) was $\theta_c=0.37$ at a temperature of $\theta=1/3$. In (d) no
  liquid-gas phase boundaries can lie outside the white area and still
  have equal pressures for the liquid and gas phase.}
\label{Fig1}
\end{figure}

Up to second order in the derivatives both of these approaches lead to
the same hydrodynamic equations
\begin{eqnarray}
\partial_t\rho+\nabla(\rho\h{u})&=&0\\
\rho \partial_t\h{u}+\rho \h{u}.\nabla\h{u}&=&-\nabla P+\nabla\sigma
\end{eqnarray}

For the van der Waals pressure for a critical density of 1 and a
temperature $\theta=\frac{1}{3}$ the pressure tensor is given by
\begin{equation}
P=\left(\frac{\rho}{3-\rho}-\frac{9}{8} \rho^2\theta_c\right)\mb{1}.
\label{VdW}
\end{equation}
Using this pressure for the two methods we observe phase-separation
below the critical temperature as shown in Figure \ref{Fig1}.  There
are, however, a number of peculiarities. 

Let us first discuss the results for the pressure method. There is no
unique solution. Three different profiles for the pressure method are
shown in Figure \ref{Fig1}(a). All of them are stable solutions which
are stable against small perturbations. Also the density profiles are
sharp, switching from the liquid to the gas density in only one
lattice spacing. These profiles correspond to a constant pressure
profile (Fig. \ref{Fig1}(b)), which is exact to machine accuracy. The
chemical potential profiles, however, are vastly different as shown in
Figure \ref{Fig1}(c). Only the middle profile corresponds to a
constant chemical potential, and therefore to the thermodynamic
equilibrium. In Figure \ref{Fig1}(d) we show the phase-diagrams. Only
in the white area in this diagram can liquid and gas phases have equal
pressures. We show the results of simulations which are initialized
with a near constant pressure in the gas and liquid phase. The $\Delta$
triangles correspond to simulations initialized with the lowest
possible constant pressure and the $\nabla$ triangles correspond to
simulations initialized with the highest possible constant
pressure. All these simulations are stable to small perturbations. It
appears that all states with equal pressures, including but not
limited to the true equilibrium state, are stable solutions.

The forcing method leads to a unique profile which has an interface
extending over several lattice spacings. The bulk pressures of the
liquid and gas phases agree, but there are variations of the pressure
(\ref{VdW}) around the interface in Figure \ref{Fig1}(b). The bulk
values of the chemical potential, however, are not equal. It is
therefore not surprising that the phase-diagram differs from the true
equilibrium profile as shown in Figure \ref{Fig1}(d).

To understand these results let us recall that expression for the
pressure (\ref{VdW}) does not contain any gradient terms. This
corresponds to $I(\nabla\rho,\nabla^2\rho,\ldots)=0$ in the expression for the free energy
(\ref{GenFree}). This means that we have not input any interfacial
free energy contributions and we therefore expect a sharp
interface. But without interfacial contributions the differential
equation $\nabla P=0$ is no longer a differential equation. Instead this
only requires that the pressure for both phases is the same,
$p_l=p_g$. Any such pressures will fulfill $\nabla P=0$, as shown in Figure
\ref{Fig1}b. In this Figure numerical solutions at the maximum and
minimum of the coexistence liquid and gas pressures as well as the
true equilibrium pressure are shown. Only for the true equilibrium
densities will $\mu_l=\mu_g$ be true as can be seen in Figure
\ref{Fig1}c. So clearly $\nabla P=\rho\nabla\mu$ is no longer generally valid. There
is no guarantee that a system without interfacial energy obeying the
dynamic equations (\ref{cont}) and (\ref{NS}) will move towards true
equilibrium. This explains why the pressure method, which leads to the
expected sharp interface, can fail to recover the equilibrium
densities.

This, however, is not enough to understand the stability of the
interfaces to small perturbations. To change the liquid and gas
densities it is in general necessary to move the interface. Because
the method leads to a sharp interface, moving this interface will lead
to states that have one lattice point with a density between the gas
and liquid densities. If the gas density is slightly increased, the
pressure will also increase, favoring a return to the original
density. Likewise if at one point the liquid density is reduced this
will lead to a lower pressure inducing the return to the original
density commensurate with the surrounding pressure. So the stability
of these non-equilibrium structures occurs because moving an interface
requires different discretizations of the interface.  And these
discretizations lead to position dependent pressures which counteract
the movement of the interface.  

The results shown in Figure \ref{Fig1} are at a mean velocity of
zero. The situation changes when a mean velocity is added to the
system. In these cases we do not find advected stationary solutions,
but instead large oscillations are observed for sharp interfaces.
This violation of Galilean invariance is removed by increasing
the width of the interface, as shown in \cite{Li}.

In the case of the forcing method the situation is more
complicated. The forcing method leads to a unique extended interface
as shown in Figure \ref{Fig1}a. Since we do not obtain a sharp
interface the pressure $P$ of equation (\ref{VdW}) is not constant, as
seen in Figure \ref{Fig1}b. Therefore higher order gradient terms must
be present in the pressure for this lattice Boltzmann method. These
gradient terms arise because of higher order terms in the lattice
Boltzmann method that were not picked up by our second order
expansion. The bulk values
of the pressure in Figure \ref{Fig1}b are the same for the
liquid and gas phases as is to be expected from the condition of
mechanical equilibrium. The same, however, is not true for the
chemical potential in Figure \ref{Fig1}c. Because the bulk values of
the chemical potential are not the same the densities also do not
correspond to their equilibrium values.

For the pressure method there are two potential remedies for us to
recover the equilibrium bulk densities for the liquid and gas
phases. The most obvious one, in light of the present discussion, is
to explicitly include the correct gradient terms in the pressure and
we will follow that route below. A second potential remedy lies in
including fluctuation terms in the Navier-Stokes
equations. Incorporating equilibrium fluctuations allows the
interfaces to move and to select the correct equilibrium bulk
densities for the pressure method. This approach, however, is outside
the scope of the current paper.

\begin{figure}
\subfigure[]{
\includegraphics[width=0.77\columnwidth]{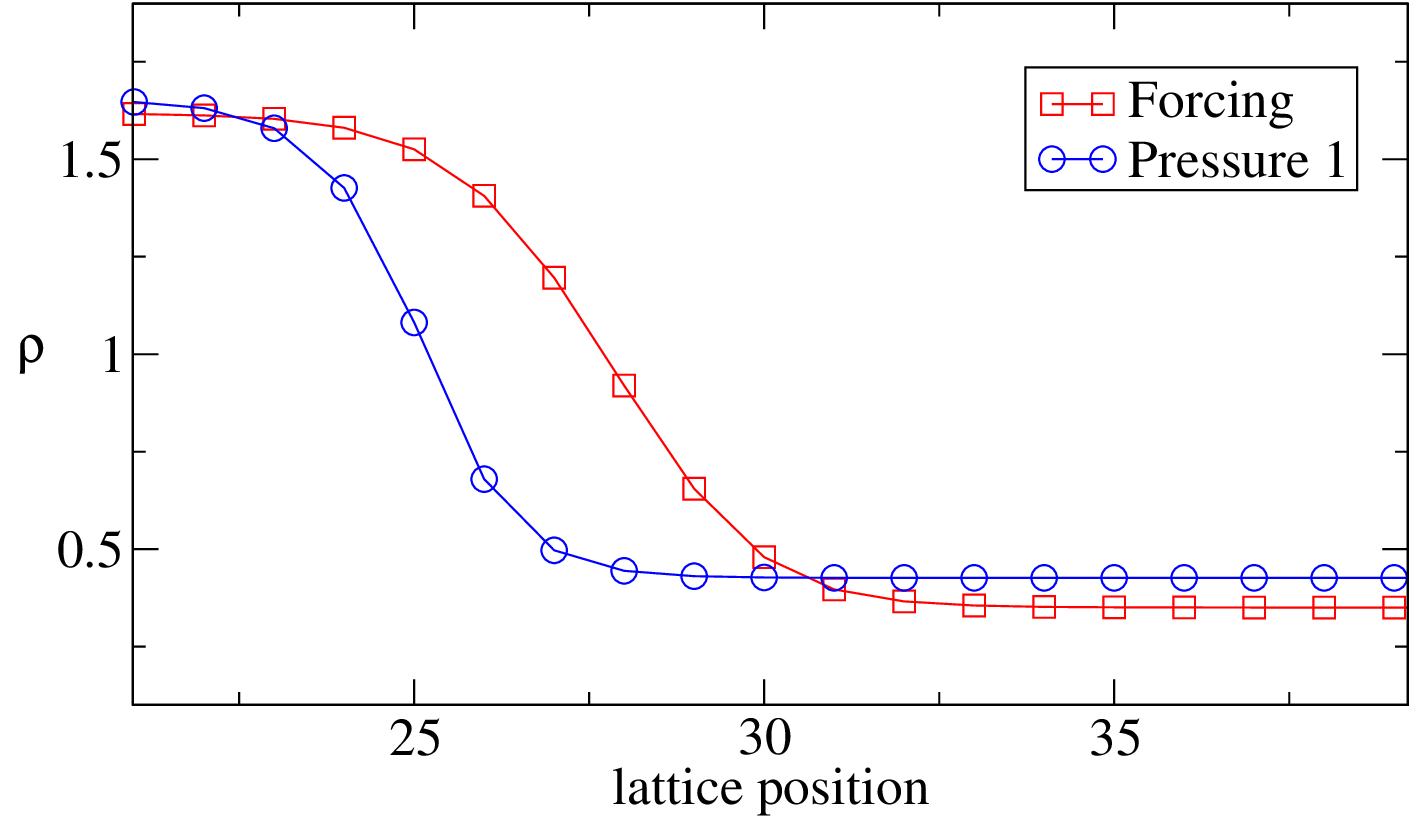}
}
\subfigure[]{
\includegraphics[width=0.77\columnwidth]{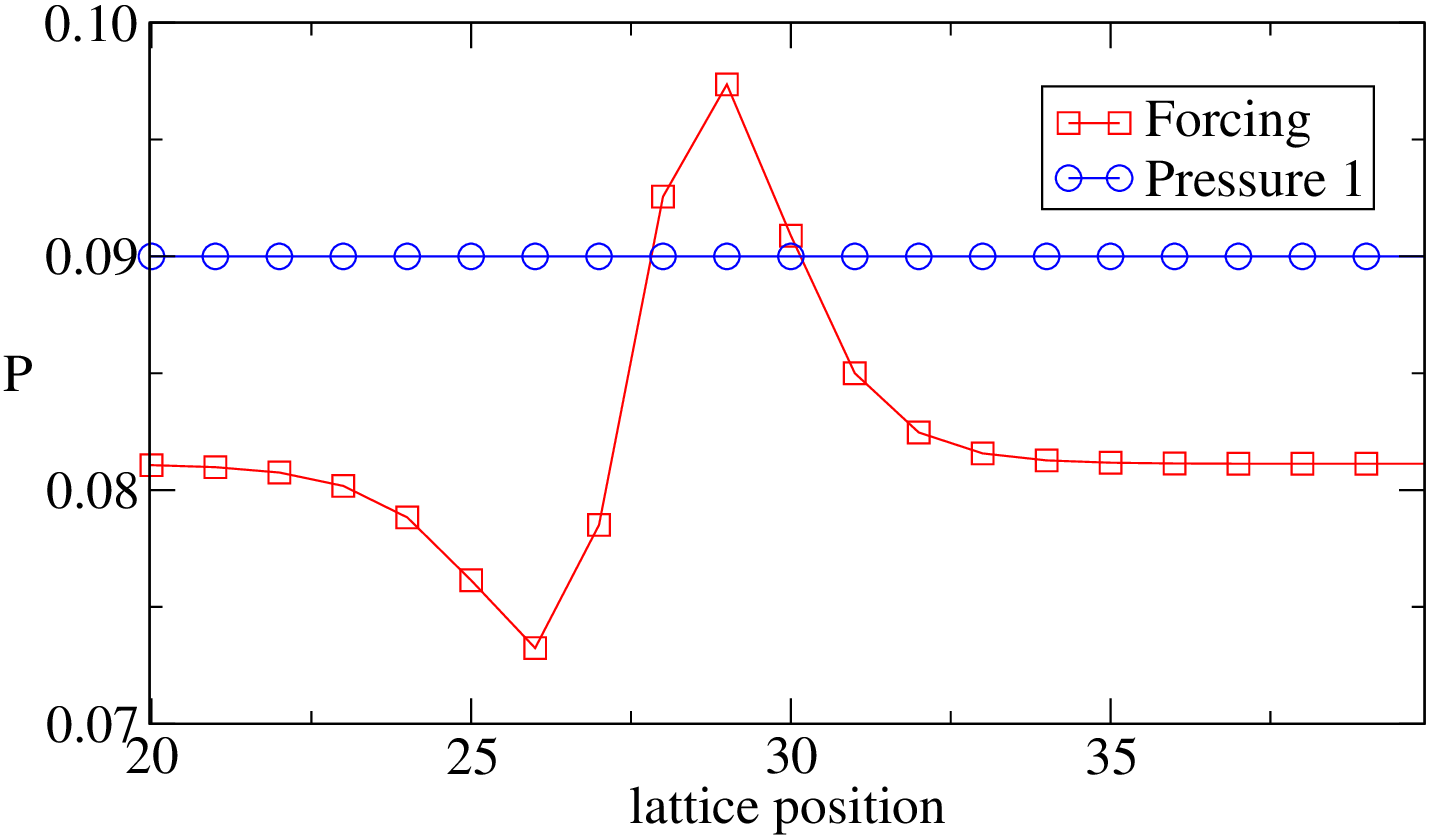}
}
\subfigure[]{
\includegraphics[width=0.77\columnwidth]{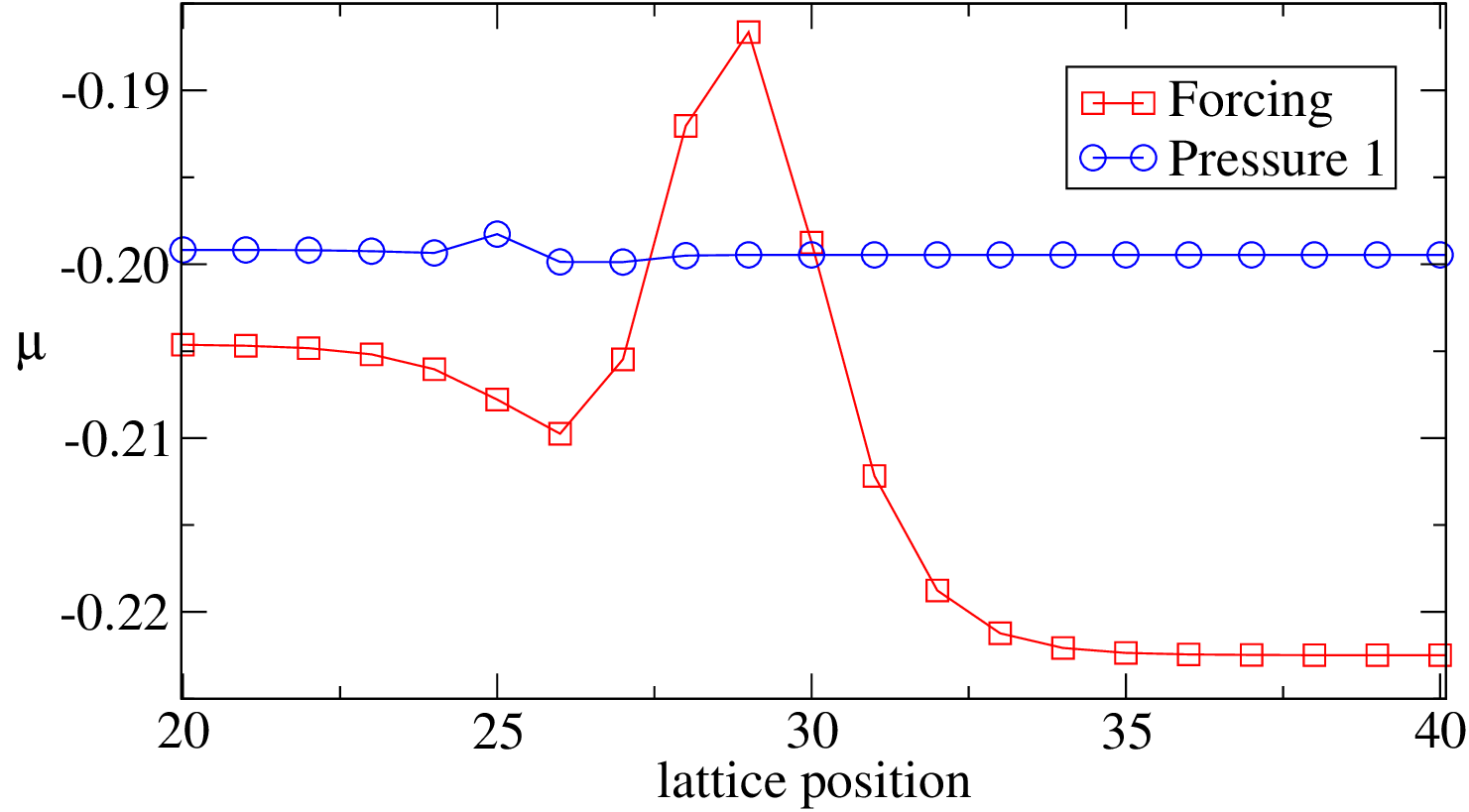}
}
\subfigure[]{
\includegraphics[width=0.77\columnwidth]{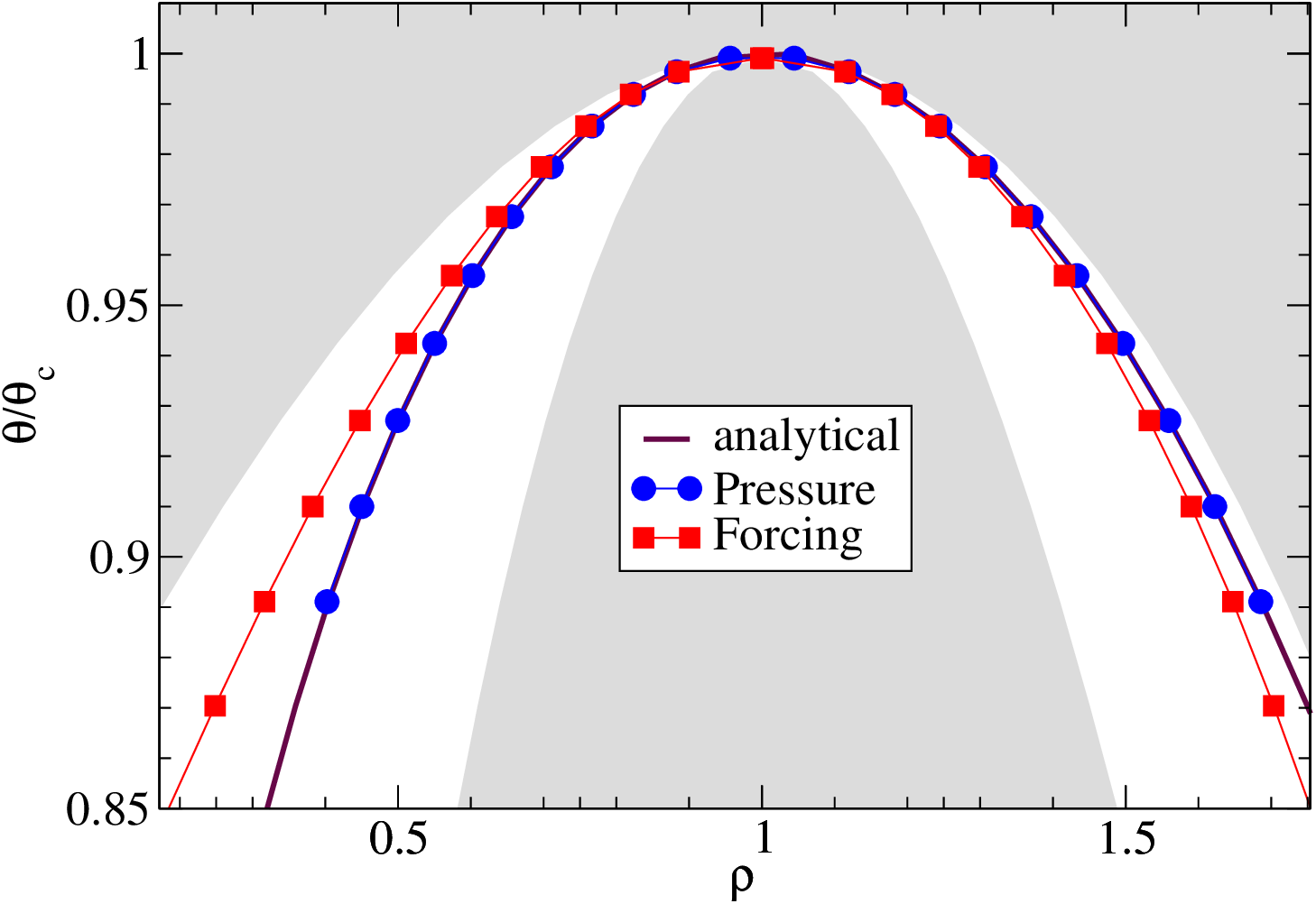}
}
\caption{The phase behavior for a liquid gas system when the explicit
  interfacial energy is included. For this set of simulations we used
  $\kappa=0.1$ in the expression for the pressure (\ref{VdW_int}). All
  other parameters are the same as for Figure \ref{Fig1}. }
\label{Fig2}
\end{figure}

While derivatives in the pressure tensor are not consistent with the
second order expansion, it appears reasonable to include the full
derivative terms in the pressure tensor. This is indeed what was done
in the original pressure method \cite{Swift} and since the numerical
simulations do not indicate the presence of any spurious interfacial
terms it is not surprising that this approach is successful for the
Pressure algorithm.

The usefulness of including the full gradient terms in the pressure
tensor is less obvious for the forcing method where there are clearly
already significant, spurious interfacial terms present that lead to an
extended interface.

So we will now replace (\ref{VdW}) with
\begin{equation}
P=\left[\frac{\rho}{3-\rho}-\frac{9}{8}
  \rho^2\theta_c-\kappa(\rho\nabla^2\rho+\frac{1}{2}\nabla\rho.\nabla\rho)\right]\mb{1}+\kappa\nabla\rho\nabla\rho.
\label{VdW_int}
\end{equation}
For a numerical implementation the gradient and Laplace operator have
to be replaced by discrete versions. This is problematic because
it will in general break the relation $\nabla P=\rho\nabla\mu$, and thereby also the
exact thermodynamic consistency. For interfaces that are wide enough
though, this relation will still hold to good approximation. We have
not yet been able to identify a discrete derivative operator and an
expression for the pressure that preserves $\nabla^D P=\rho\nabla^D\mu$. Such an
expression could guarantee that a constant pressure is exactly
equivalent to a constant chemical potential, and therefore it would
exactly recover equilibrium thermodynamics. 

For the simulations presented in this paper we require the one
dimensional discrete gradient and Laplace operator. We use the
discretization
\begin{eqnarray}
\nabla^{(D)} f(x)=\frac{1}{2}(f(x+1)-f(x-1)),\\
\nabla^{2(D)} f(x)=f(x+1)-2f(x)+f(x-1).
\end{eqnarray}
For $\kappa=0$ the expression (\ref{VdW_int}) leads to (\ref{VdW}). But for
finite $\kappa$ we now expect to find an extended interface for the
pressure method. We also expect that a constant pressure will now
reduce the difference in the chemical potential for the two
phases. The results are shown in Figure \ref{Fig2}.

For a sufficiently large interfacial energy contribution of $\kappa=0.1$
there is now a unique solution for the pressure method. This solution
also agrees very well with the analytical phase-diagram. We find that
the pressure is constant up to machine accuracy and the chemical
potential is very nearly constant. In particular the deviation in the
bulk value of the chemical potential in the gas and liquid phase are
less than $3\times 10^{-4}$.

\begin{figure}
\begin{center}
\includegraphics[width=0.8\columnwidth]{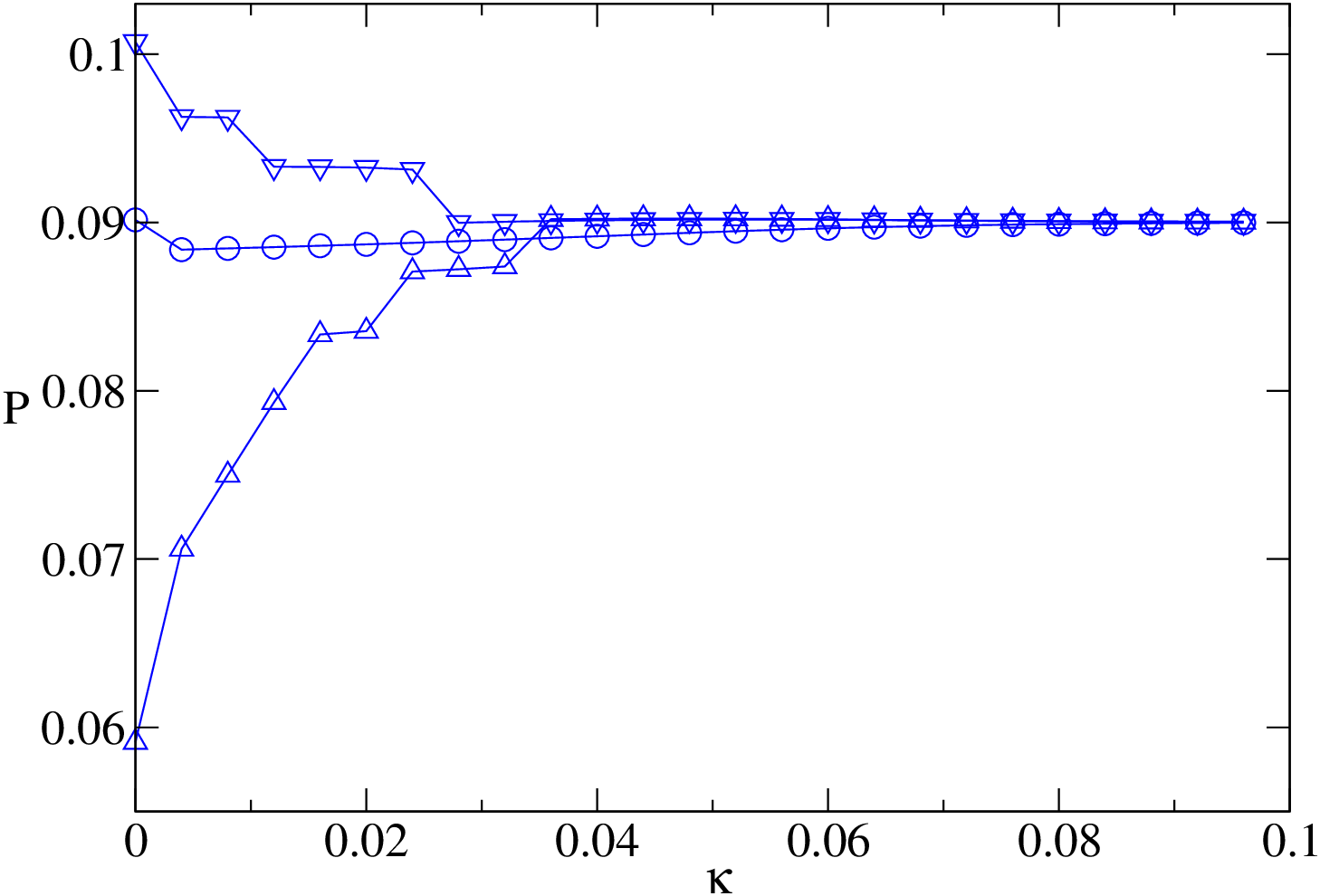}
\end{center}
\caption{Convergence of the pressure in the two-phase system for
  different initial conditions for $\theta=0.37$. A unique solution is only
  found for large enough values of $\kappa$. For a detailed discussion of
  this effect see the main text.}
\label{Fig3}
\end{figure}

In the continuous situation the non-definiteness of the stationary
state is limited strictly to $\kappa=0$. For a discrete approach
$\nabla^DP=\rho\nabla^D\mu$ is not generally valid, and therefore it is not \textit{apriory}
clear that an arbitrarily small interfacial term in the pressure will
guarantee a convergence to the equilibrium solution. As we have seen
before it is not even guaranteed that it will converge to a unique
solution.  We therefore performed a set of simulations for different
values of $\kappa$ from 0 to 0.1 with initial conditions that correspond to
the extreme values of the equal pressure as well as the equilibrium
solution. The simulation results are shown in Figure \ref{Fig3}. We
see that non-uniqueness survives for finite $\kappa$. The range of possible
pressures is rapidly reduced until a near-unique solution is found at
$\kappa$ larger than about $0.08$. 

The origin of this non-uniqueness even in the presence of an
interfacial energy lies in the discretization of the interface. Moving
the interface over the lattice requires a re-discretization of the
interface. Each of these discretizations will have a different
interfacial free energy $\sum_i (\kappa/2)(\nabla \rho)^2$. To visualize this imagine
a sharp interface of Fig \ref{Fig1}a. This interface has only two
points with non-vanishing gradient and Laplace operators. If this
interface moves there will have to be at least one point with
intermediate value of $\rho$ and now there are at least three points
with non-vanishing gradient and Laplace terms. Therefore
moving the interface will change the total free energy.
Mass-conservation generally implies that changing the liquid and gas
densities requires changing the relative gas and liquid
volumina. Therefore changing the densities requires that the
interfaces move. But if moving the interface requires passing a local
maximum in the free energy, this implies that the system is at least in
a metastable state. And a lattice Boltzmann method without
fluctuations cannot escape such a metastable state.  So to move an
interface that is well aligned with the lattice one has to overcome a
free energy barrier with respect to the less favorable discretization
of the interface and the simulation can get stuck in a non-equilibrium
configuration.

The Forcing method is also noticeably improved through the inclusion
of the gradient terms. The difference in the bulk chemical
potential is about halved from $0.03$ to $0.015$. This is still about
50 times larger than the difference for the pressure method. This
difference in the chemical potential translates to a deviation of the
liquid and gas densities from their equilibrium values. This can be
clearly seen in Figure \ref{Fig2}d. In order to understand the
origin of the large deviations of the forcing method from the
thermodynamic equilibrium we need to uncover the spurious gradient terms in
the effective pressure. To do this we need to improve our expansion method to
consistently derive those derivative terms.

\section{Higher order near equilibrium expansion}
To identify the higher order terms of the equilibrium structure we
perform a 5th order expansion around an equilibrium profile which is
stationary, but possibly advected with a constant velocity $U$. As
before we make the assumption that derivatives are small
($O(\partial)=O(\epsilon)$). From equation (\ref{NSeqn}) we already know that
$O(F)=O(\partial)=O(\epsilon)$. To avoid obtaining too many terms we will limit our
analysis here to small velocities, so we also postulate
$O(U)=O(\epsilon)$. Note that since we will only neglect terms of order
$O(\epsilon^5)$ we will not loose any of the terms present in the lower order
expansion, with the single exception of the $\nabla^2Q$ term in
equation (\ref{sigma}).

Performing a Taylor expansion to 5th order in the derivatives of
(\ref{LBeqn}) and expressing the $f_i$ in terms of $\hat{f}_i^0=f_i^0+A_i$ we obtain
\begin{eqnarray}
&&F_i+D(\hat{f}_i^o-\tau F_i)-(\tau-\frac{1}{2}) D^2(\hat{f}_i^0-\tau
  F_i)\nonumber\\
&&+(\tau^2-\tau+\frac{1}{6})D^3(\hat{f}_i^0-\tau F_i)\nonumber\\
&&-(\tau^3-\frac{3}{2}\tau^2+\frac{7}{12}\tau-\frac{1}{24})D^4\hat{f}_i^0 
+O(\epsilon^5)\nonumber\\
&=&\frac{1}{\tau} (\hat{f}_i^0-f_i)
\label{LBlong}
\end{eqnarray}
which is the extension of (\ref{expand_equil}) to two more
orders. Here the derivative operator is
\begin{equation}
D\equiv\partial_t + v_i.\nabla.
\end{equation}

For a stationary profile, advected with velocity $U$, we have the
operator identity
\begin{equation}
\partial_t+U.\nabla=0.
\end{equation}
This simplifies the derivative operator
\begin{equation}
D= (v_i-U).\nabla.
\end{equation}
We now need to take the zeroth and first order moments of this
expression to obtain the expressions for the continuity and Navier
Stokes level equations in the stationary advected limit. The
expectation is that the continuity equation takes the form
\begin{equation}
\nabla(\rho \hat{u}-\rho U)=O(\epsilon^5)
\label{contStat}
\end{equation}
and the Navier-Stokes level equation will become
\begin{equation}
\nabla(P)=O(\epsilon^5).
\label{NSStat}
\end{equation}
This then allows us to identify the effective mean fluid velocity
$\hat{u}$ to fifth order and the effective pressure tensor $P$ also to
fifth order. It is highly desirable, although far from obvious, that
these quantities be independent of the relaxation time
$\tau$. Otherwise the equilibrium properties would be coupled to the
transport coefficients.

This task is cumbersome since it involves up to fifth order velocity
moments of the equilibrium distribution function. We will restrict
ourselves here to the analysis of a projection of the most common
models (D2Q7,D2Q9,D3Q15,D3Q19,and D3Q27) to one dimension. This
projection is the D1Q3 model, a one dimensional model with three
velocities. It is important to note that this analysis is entirely
sufficient to determine the phase-behavior of all the above
models. The only issue not addressed by this analysis is the isotropy
of the models. One caveat is that this analysis will only be able to
make statements of the equilibrium behavior of the method, not about
the approach to equilibrium.

\subsection{Analysis of the D1Q3 model}
The D1Q3 model is a one dimensional lattice Boltzmann model with three
velocities: $v_i=\{-1,0,1\}$.  In this model there are only three
distinct velocity moments at each lattice point, corresponding to the
three densities $f_i$. The moments of the equilibrium density are
\begin{eqnarray}
&\sum_i (\hat{f}_i^0+A_i) = \rho, \;\;\; \sum_i (\hat{f}_i^0+A_i) v_i=\rho u,\nonumber\\
&\sum_i \hat{f}_i^0v_i^2=\rho uu+\frac{\rho}{3}+A.
\end{eqnarray}
Because
$v_i^3=v_i,\;v_i^4=v_i^2$ \textit{etc.}, all higher order velocity moments are given
by these first three moments. The three moments of the forcing term
are given by
\begin{equation}
\sum_i F_i=0, \;\;\;\sum_i F_i v_i= F \;\;\; \sum_i F_i v_i^2= 2 Fu+\Psi.
\end{equation}
The higher order moments are similarly given by these first three
moments.

The algebra involved in calculating the higher moments is quite
extensive, so I will only outline the method here without giving the
lengthy intermediate results. Summing $\sum_i (\ref{LBlong})$ gives an
expression for the continuity involving all moments of $\hat{f}_i^0$
and $F_i$. Similarly we obtain such an expression containing all
moments by summing $\sum_i (\ref{LBlong}) v_i$ to obtain the momentum
equation. We can then use the momentum equation to express $A$ in
terms of the other moments and higher order derivatives of $A$. Then
we insert this expression repeatedly into itself (similarly to what
we did for equation (\ref{fzero_eqn})) to express $A$ completely in
terms of the other moments and their derivatives.

This expression for $A$ is then used to eliminate any $A$ dependence
in the continuity equation. We then use the resulting continuity
equation to express $u$ in terms of the other moments and higher order
derivatives of $u$. This expression for $u$ is then inserted for
all but the lowest order derivatives of $u$ of the continuity
equation. The resulting continuity equation has the form
\begin{equation}
\nabla(\rho u-\frac{1}{2} F-\rho U)-\frac{1}{4}\nabla\nabla[ F U+\frac{1}{3}\nabla(\rho U)]=0(\epsilon^5)
\label{cont3}
\end{equation}
which only contains one term of $u$ and no terms with $A$. We use this
version of the continuity equation to express $u$ in terms of $U$, $\rho$
and $F$ and replace all occurrences of $u$ in the momentum
equation. We then remove all but the lowest order occurrences of $A$
in the momentum equation and obtain
\begin{eqnarray}
& F+\nabla \left[\frac{\rho}{3}+A
-(\tau-\frac{1}{2})[3 FU+\nabla(\rho U)]
\right.\nonumber\\
&\left.+\frac{1}{4\rho} FF+\frac{1}{12}\nabla F
-\tau(\frac{FF}{\rho}+\Psi)
 \right]=O(\epsilon^5).
\label{finalNS}
\end{eqnarray}
Now we apply these results to our Pressure and Forcing
methods. 

\subsection{Pressure Method}
For the pressure method we use the moments defined in
equation (\ref{pressure1}) and obtain for the continuity equation
\begin{equation}
\nabla(\rho u-\rho U)-\nabla\nabla[\frac{1}{12}\nabla(\rho U)]=0(\epsilon^5).
\end{equation}
Comparing this to (\ref{contStat}) we see that the mean fluid velocity
is
\begin{equation}
\hat{u}=\frac{u}{1+\frac{1}{12}\;\frac{\nabla^2\rho}{\rho}}.
\label{Ucorrection}
\end{equation}
The gradient term is a new correction for the measurement of the
velocity for the pressure method. 
For the momentum equation we obtain
\begin{equation}
\nabla P=O(\epsilon^5).
\end{equation}
There are no additional pressure terms for the pressure method,
which is consistent with the fact that we found a constant input
pressure for simulations with the pressure method in Figures
\ref{Fig1}(b) and \ref{Fig2}(b).

\begin{figure}
\begin{center}
\subfigure[]{
\includegraphics[width=0.8\columnwidth]{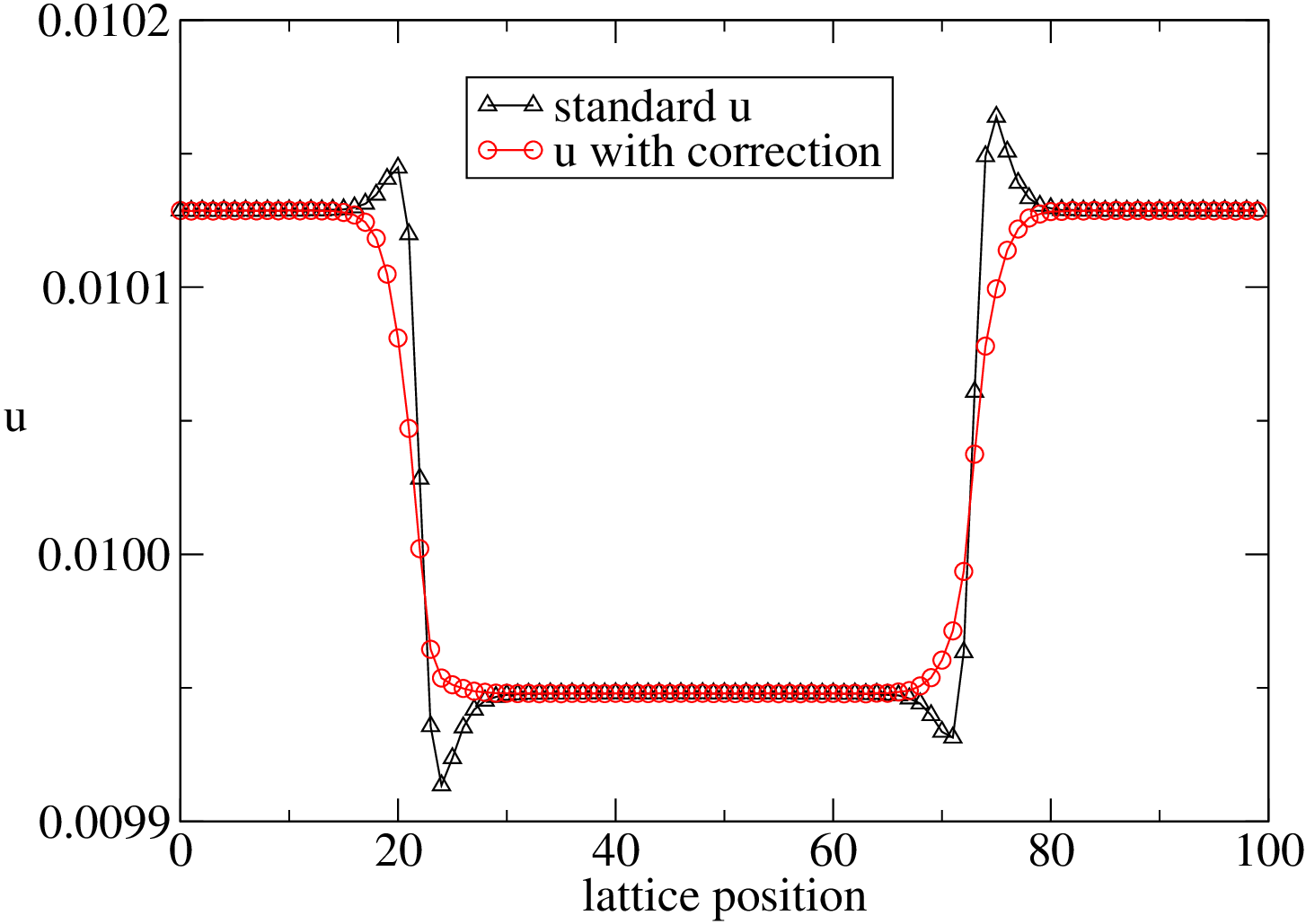}
}
\subfigure[]{
\includegraphics[width=0.8\columnwidth]{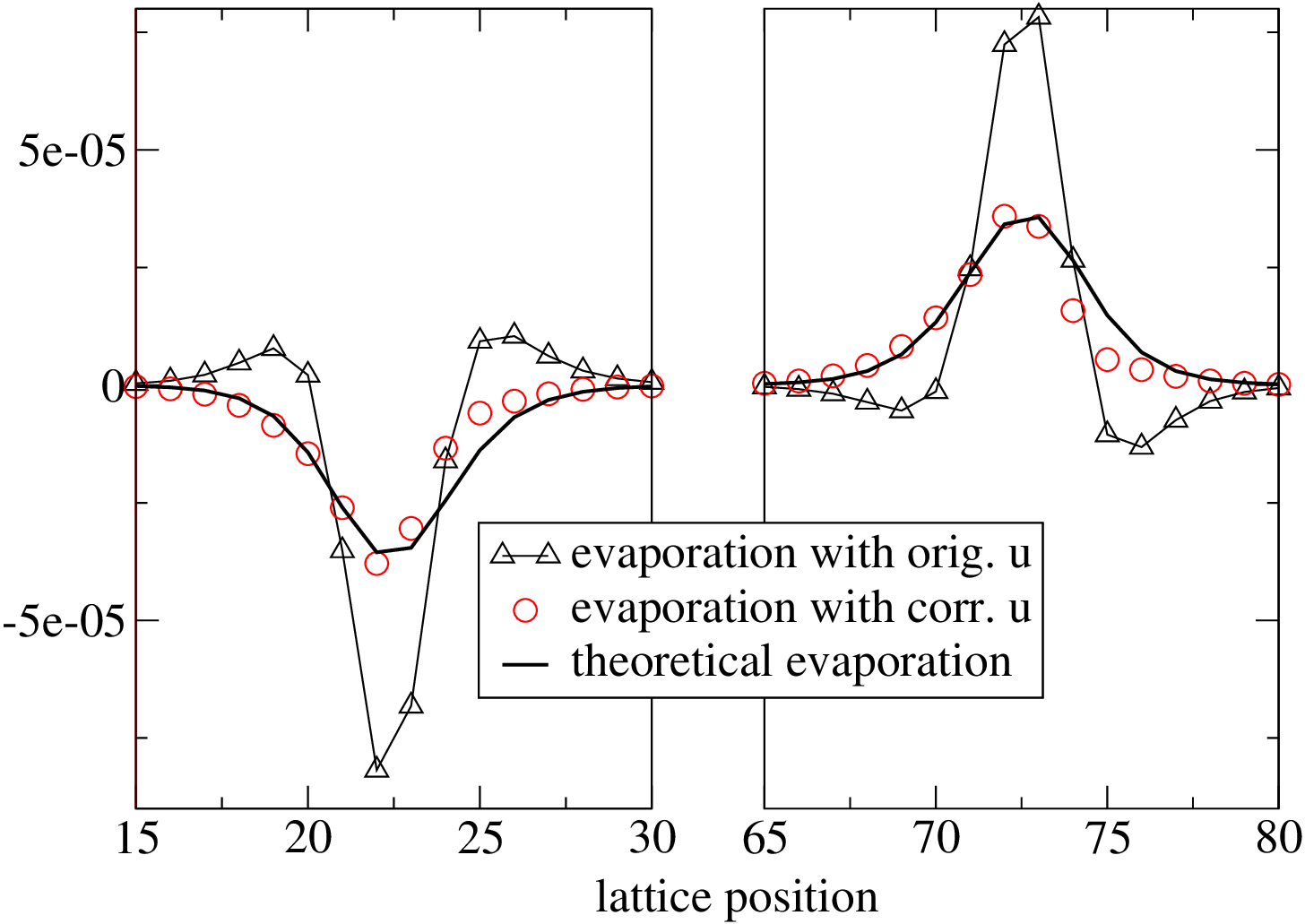}
}
\end{center}
\caption{Comparison of the expression for the mean fluid velocity from
original definition (\ref{Udef}) and the corrected version
(\ref{Ucorrection}) for the pressure method in (a). The system was
initialized with a constant velocity of $U=0.01$. While the correction
term clearly removes some spurious velocity terms there is a more
noticeable violation of Galilean invariance which advects the gas
phase faster than the liquid phase with respect to the lattice. In (b)
we compare the measured expression for the evaporation $\nabla [\rho
(U-\hat{u})]$ for the original and corrected expressions for $\hat{u}$
and compare it with the theoretical one of $\xi\nabla\rho$ for $\xi=-0.0002$.
This simulation was performed with $\theta_c=0.35$, $\theta=1/3$ and $\kappa=0.1$.}
\label{Ugraph}
\end{figure}

To test the correction for the velocity predicted in
(\ref{Ucorrection}) we have to consider a situation in which $U$ is
not zero. We set up a profile that is initiated with a velocity
of $U=0.01$. In Figure \ref{Ugraph}(a) we plot both the standard velocity
$u$ from (\ref{Udef}) and our new approximation of the true fluid
velocity $\hat{u}$ from (\ref{Ucorrection}). First we note that the
gas phase is advected faster than the liquid phase, which is a problem
of Galilean invariance \cite{Li}. The gas is advected faster than the
liquid and a constant evaporation on the leading edge of the droplet
and a condensation at the trailing interface of the liquid leads to a
mean interface velocity that is less than the imposed mean fluid
velocity of $0.01$. This evaporation and condensation leads to an
additional velocity $\xi$ of the interface. For such an interface 
velocity we have an additional contribution to the change of the
density, the rate of evaporation given by
\begin{equation}
(\partial_t+U.\nabla) \rho = -\xi.\nabla\rho = \nabla [\rho (U-\hat{u})].
\end{equation}
So to compare the original definition of $\hat{u}$ from
equation (\ref{uhatdef}) with the corrected definition of
equation (\ref{Ucorrection}) we plot $\nabla[\rho (\hat{u}-U]$ for the two
definitions of $\hat{u}$ and $\xi\nabla\rho$ in Figure \ref{Ugraph}(b). We see
that the corrected version of the velocity allows a very good fit with
the theoretical expression. The best fit for $\xi$ is -0.0002 which
corresponds to a 2\% correction for the observed domain velocity.

While the Galilean invariance violations are small, in this case it is
still worth while to correct the Galilean invariance violations of the
lattice Boltzmann method. To do that we need an expansion that does
not make the assumption $u=O(\epsilon)$. This investigation of higher order
Galilean invariance violations will be reported elsewhere.

\subsection{Forcing Method}
For the forcing method (\ref{forcing1}) we obtain the continuity
equation
\begin{equation}
\nabla \left(\rho u -\frac{1}{2}F-U\right) = O(\epsilon^5).
\end{equation}
This implies that the mean fluid velocity is given by $\hat{u}=\rho
u-F/2$ in agreement with the lower order expression
(\ref{uhatdef}). There are no further corrections to the velocity for
the forcing method to this level of approximation.

\begin{figure}
\includegraphics[width=0.9\columnwidth]{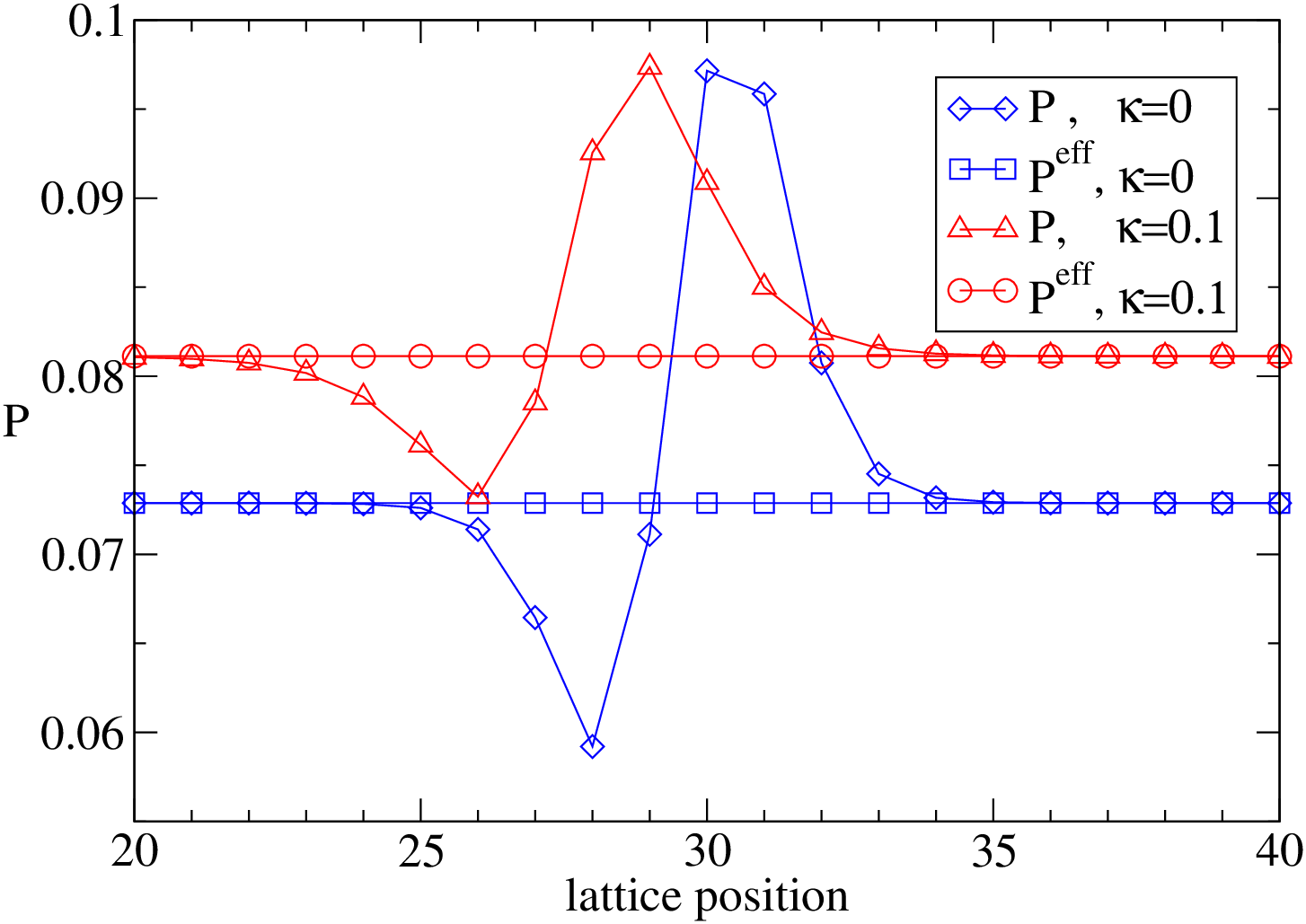}
\caption{Comparison of the input pressure P from equation (\ref{VdW_int})
  and the effective pressure $P^{eff}$ of equation (\ref{peff}) for the
  forcing method for two values of $\kappa$. We see that the effective
  pressure $P^{eff}$ is constant to machine accuracy in both
  cases. This agreement is better than expected since expansion
  predicted this pressure only up to fifth order. The critical
  temperature for this simulation is $\theta_c=0.37$ at $\theta=1/3$ so that
  true equilibrium pressure is 0.09.}
\label{Pgraph}
\end{figure}

To evaluate the higher order terms for the pressure in
equation (\ref{finalNS}) we need to insert the specific form of the
force. We have
\begin{equation}
\rho F=\nabla^DP^{nid}=\sum_{k=0}^\infty \frac{\nabla^{2k+1}}{(2k+1)!}P^{nid}.
\end{equation} 
With this expression for $F$ the momentum equation becomes
\begin{equation}
\nabla \left(P-(\tau-\frac{1}{4}) \frac{FF}{\rho}+\frac{1}{4}\nabla^2 F\right) = O(\epsilon^5),
\end{equation}
where we have again expressed some of the higher order contributions
in terms of the force.
We conclude that the effective pressure is given by
\begin{equation}
P^{eff}=P-(\tau-\frac{1}{4})\frac{FF}{\rho} + \frac{1}{4}\nabla^2F+ O(\epsilon^4).
\label{peff}
\end{equation}
This explains why the input pressure is not constant for the current
forcing method. To verify this analytical result we can plot both the
input pressure $P$ and the predicted effective pressure $P^{eff}$ for
a phase separated stationary profile obtained from the forcing version
of the lattice Boltzmann simulation. The result is shown in Figure
\ref{Pgraph}.

Somewhat surprisingly, the pressure $P^{eff}$ is not simply a fifth
order approximation to the true pressure. Instead, for $U=0$, this
gives a pressure that is constant to machine accuracy! To be exact
what is constant is the discretization
\begin{equation}
P^{eff}=P-(\tau-\frac{1}{4})\frac{FF}{\rho}+\frac{1}{4} \nabla^{2(D)}
P-\frac{1}{12} \nabla^{2(D)} \rho.
\label{Peff}
\end{equation}
This is, presumably, due to the fact that the higher order moments are
just repeats of the lower order moments and this allows the higher
order moments to consist of higher order derivatives of the lower
order moments. Unfortunately this exact solution no longer holds if
$U\neq0$, because the higher order moments contain higher powers of
$U$. The fact that we know an \textit{exact} pressure that we minimize
may prove to be very useful for diffusive systems, though, for which we
have $U=0$.

This may also open the door for an important application of mimetic
calculus. If we were able to find some discrete gradient operators
$\nabla^\delta$ for which $\nabla^\delta P=\rho \nabla^\delta \mu$ holds, we would be able to devise a
method that \textit{always} recovers the correct equilibrium behavior.

\subsection{A new forcing method}

\begin{figure}
\begin{center}
\includegraphics[width=0.9\columnwidth]{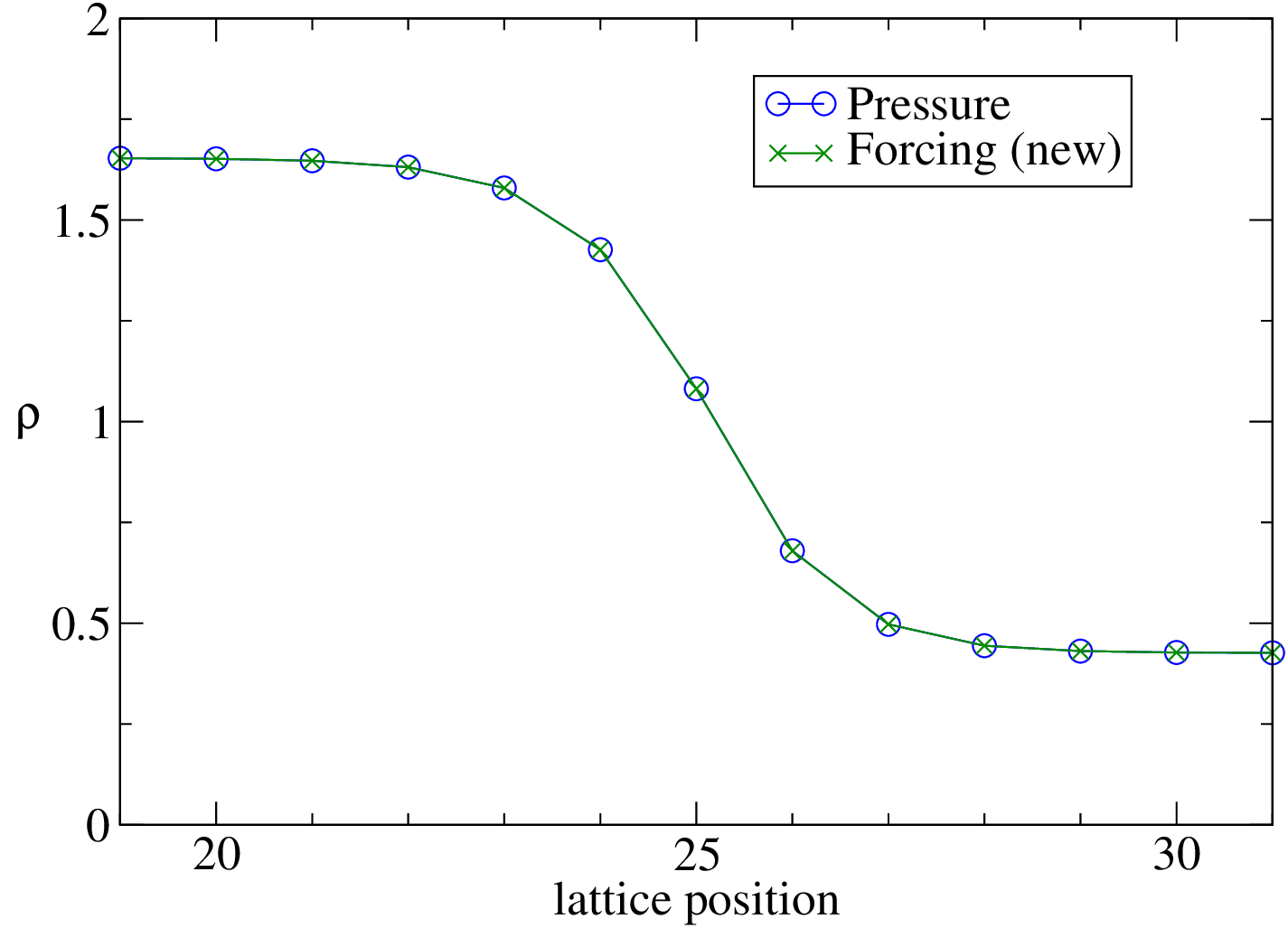}
\end{center}
\caption{The performance of the corrected Forcing method gives results
  that are identical to the results of the pressure method. The
  interfaces correspond exactly. Therefore the pressures and the
  chemical potentials also agree. Simulation parameters are
  $\theta_c=0.37$, $\theta=1/3$ and $\kappa=0.1$.}
\label{ncomp}
\end{figure}

\begin{figure}
\subfigure[]{
\includegraphics[width=0.9\columnwidth]{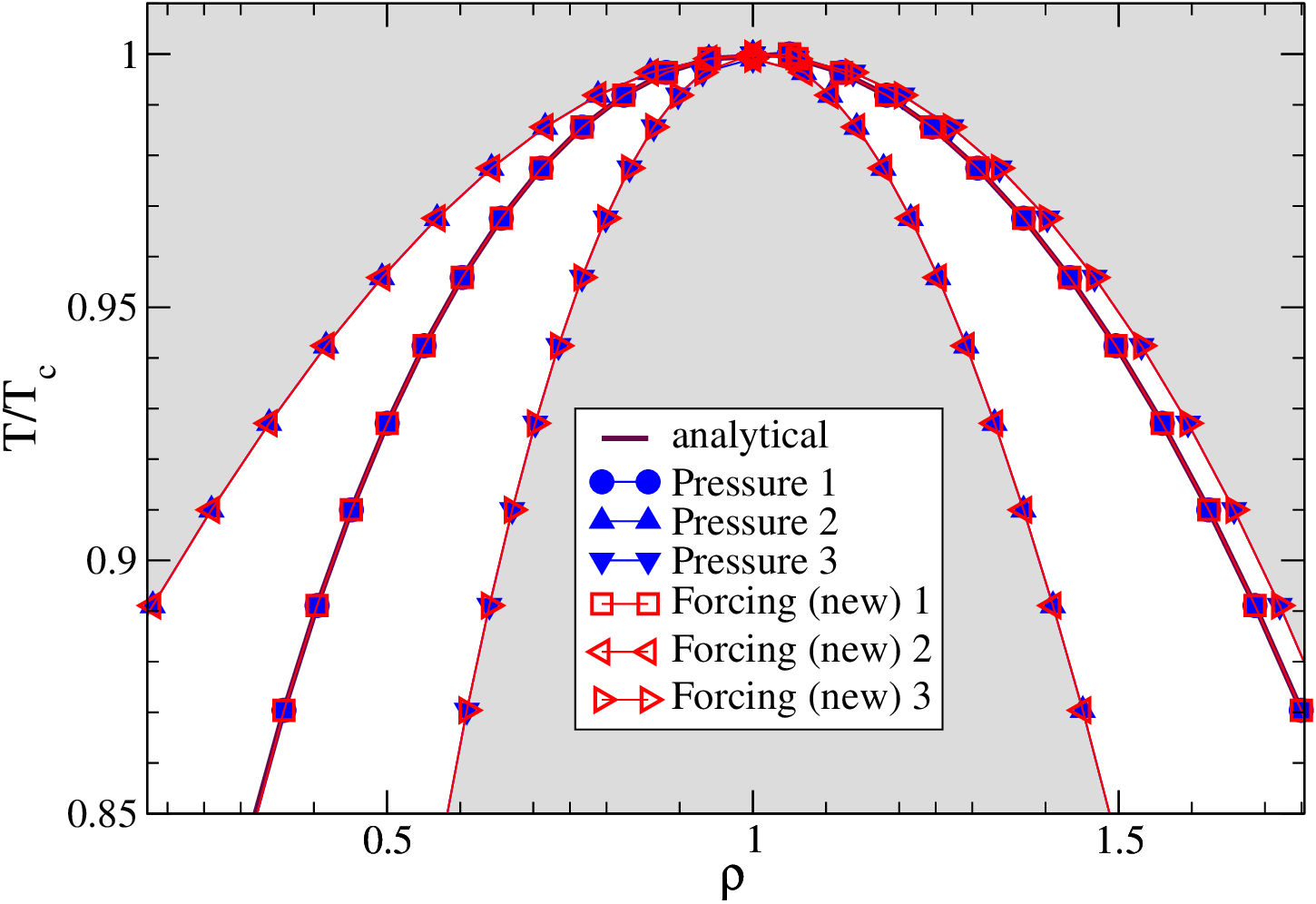}
}
\subfigure[]{
\includegraphics[width=0.9\columnwidth]{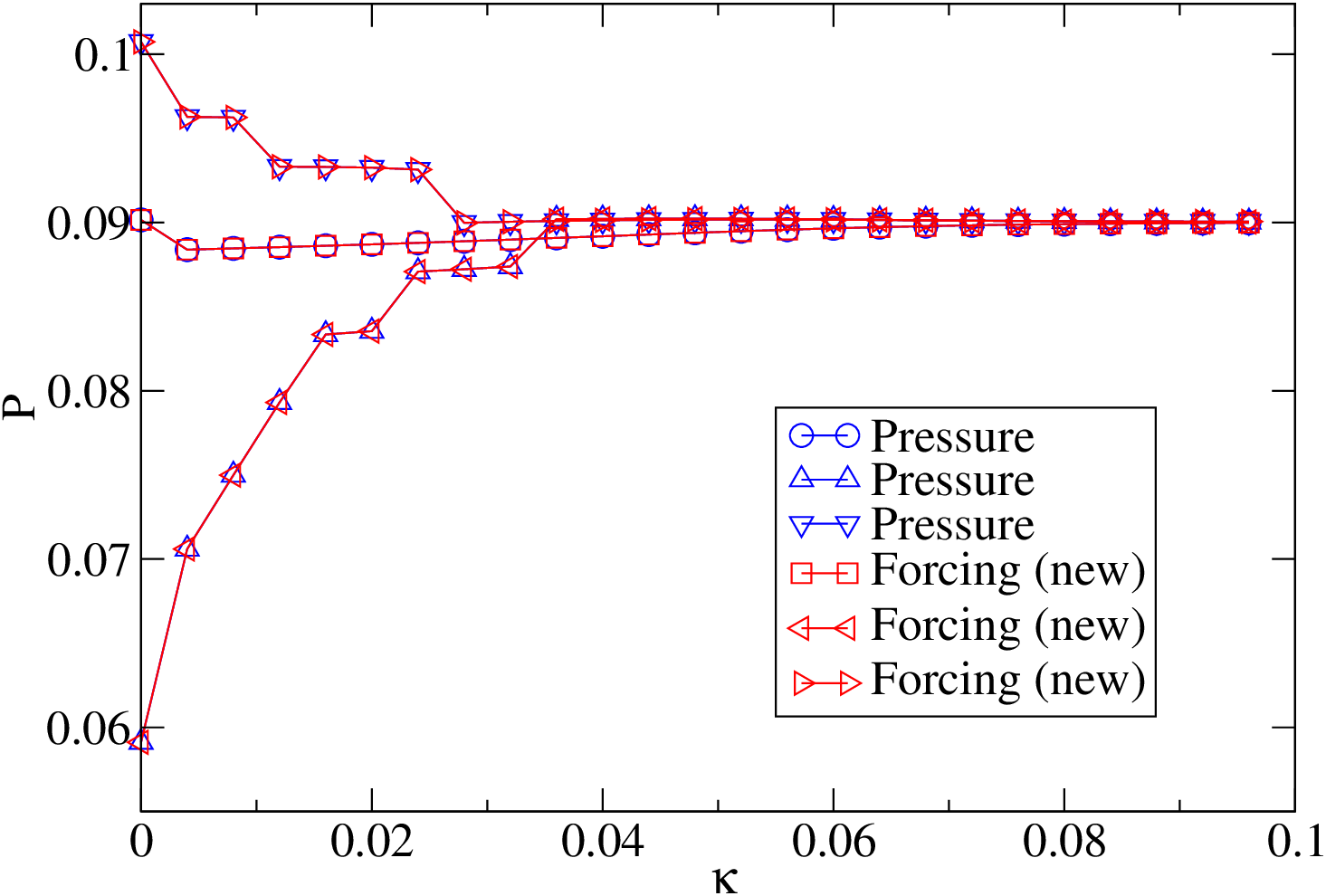}
}
\caption{The pressure and the new forcing methods give
  equivalent results. We see that the phase-diagram for $\kappa=0$ even reproduces
  the non-uniqueness of the solutions. The recovery of a universal
  profile also occurs at the same pace in the new forcing and pressure
  methods, as shown in (b).
}
\label{FigComp}
\end{figure}

Now that we have identified the effective pressure $P^{eff}$ that is
constant in the steady state we can identify a method that will have a
constant input pressure $P$. Investigating equations (\ref{cont3}) and
(\ref{finalNS}) we can identify a choice of $\Psi$ that will allow us to
cancel the additional pressure terms due to the force. We can amend the
original forcing method (\ref{forcing1}) with
\begin{equation}
\tau\Psi=(\tau-\frac{1}{4})\frac{FF}{\rho}+\frac{1}{12}(\nabla\nabla)^D\rho.
\label{forcing2}
\end{equation}
We then obtain the continuity equation 
\begin{equation}
\nabla\left(\rho u-\frac{1}{2} F-U\right) = O(\epsilon^5)
\end{equation}
and the momentum equation
\begin{equation}
\nabla P= O(\epsilon^5).
\end{equation}
The choice of $\Psi$ is not unique regarding the exact choice of the
correction terms. We could replace $(\nabla\nabla)^D\rho$ with $-3\nabla^DF$ and still
remain correct up to $O(\epsilon^5)$. However,
the above choice again leads to a pressure that is constant up to machine
accuracy for $U=0$, just as we did with the pressure method.

A comparison of the pressure method and the corrected forcing method
is shown in Figure \ref{ncomp}. The new forcing method is nearly
indistinguishable from the pressure method. One small difference
between the two methods is that an alternating pattern in the pressure
does not lead to a force, and therefore does not decay. In the
simulation shown in Figure \ref{ncomp} those pressure oscillations
have an amplitude of about $10^{-5}$. However, in some simulations
these oscillations can become large. Also for large velocities these
oscillations can become unstable and make this method less stable than
the pressure method at larger velocities.

\begin{figure}
\includegraphics[width=0.9\columnwidth]{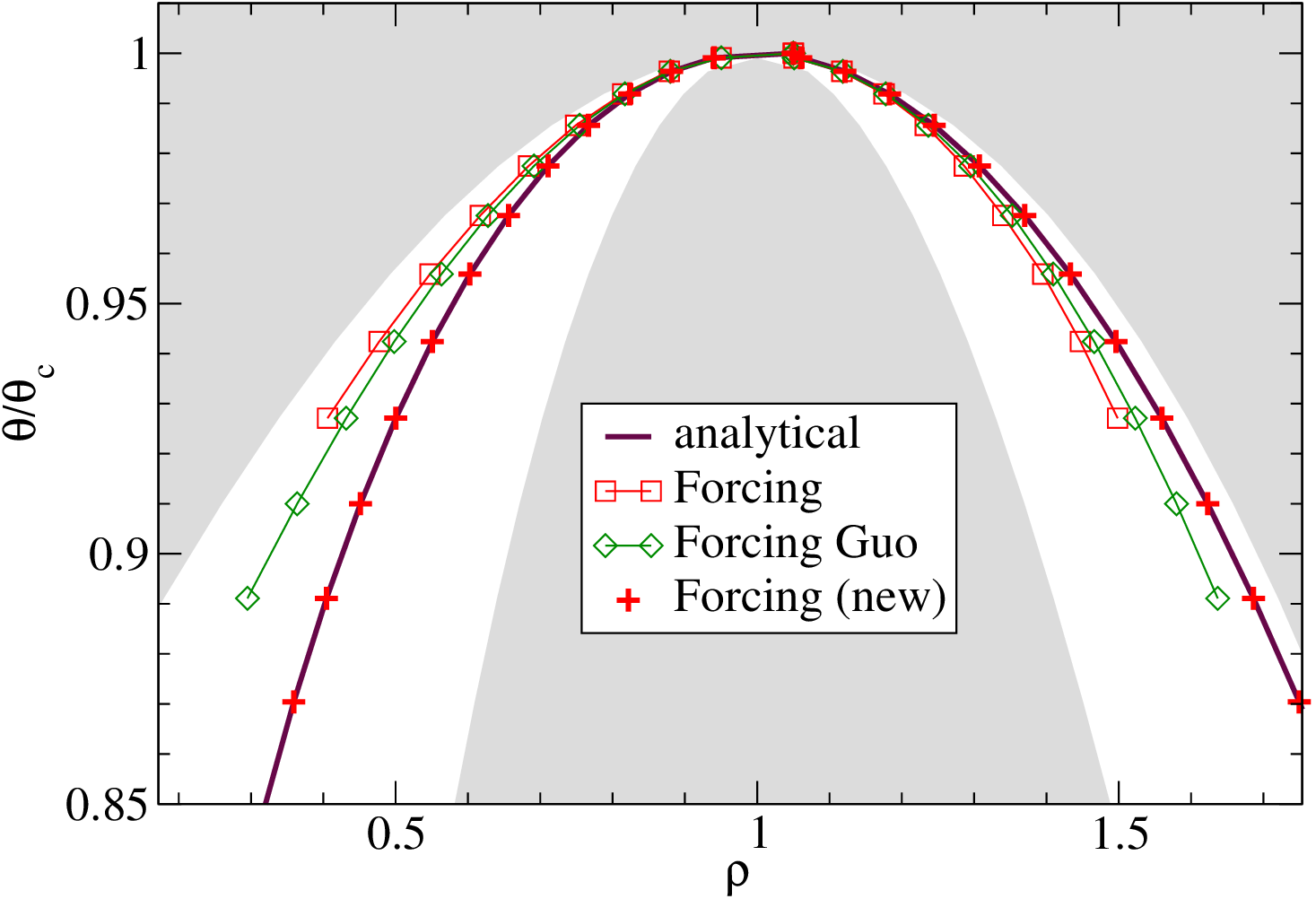}
\caption{Comparison of the effect of correction terms to the forcing
  method derived by Guo \textit{et al.} \cite{Guo} and the current
  paper. We use the pressure of equation (\ref{VdW}). We observe that the
  agreement with the phase-diagram is improved using the correction
  terms from \cite{Guo}. For the new forcing method the results would
  be indeterminate just at the results of the pressure method in
  Figure \ref{Fig1}. However, including gradient terms of
  (\ref{VdW_int}) with $\kappa=0.1$ we recover thermodynamic consistency.}
\label{FigGuo}
\end{figure}

Because the pressure and chemical potential depend only on the profile
$\rho(x)$ the new forcing method gives the same constant pressure and
near constant chemical potential as the pressure method shown in Figure
\ref{Fig2}.  This forcing method is therefore thermodynamically
consistent. It also recovers the analytical phase diagram, as shown in
Figure \ref{FigGuo}.

We also examined the behavior of the forcing method in the limit where
$\kappa\to0$. This limit may appear tricky because the correction terms need
to cancel the numerical derivatives we uncovered in the
expansion. This should uncover any higher order terms that we missed
in our expansion. We find, however, that the method described here
will even recover the sharp interfaces we observed for the pressure
method. The non-uniqueness of the solutions which we found for the
pressure method is also observed for the pressure method.
The recovery of a unique solution for larger values of $\kappa$ occurs at
the same pace as for the pressure method. This is shown in Figure
\ref{FigComp}. This surprising result is due to the fact that the
pressure $P$ is exactly constant for both methods.

\section{Discussion}
In this paper we have shown that forcing terms lead to non-negligible
higher order terms for systems with large density gradients. We
introduced a new equilibrium analysis method, that allowed us to
identify correction terms up to 5th order. This equilibrium analysis
allowed us to identified those correction terms. This analysis
analysis gave the exact pressure for systems that are not advected
with respect to the lattice.

Correction terms similar to, but different from, the correction terms
we identified have been proposed by Guo \textit{et al.}
\cite{Guo}. Their analysis leads them to conclude that for a body
force we should choose (in the notation of this paper)
\begin{equation}
\tau \Psi=\left(\tau-\frac{1}{2}\right) \rho FF.
\label{eqnGuo}
\end{equation}
Comparing this to (\ref{forcing2}) we find that this term is
different in that we predict a factor of $1/4$ instead of $1/2$ as
well as an additional double derivative term of $\rho$. The correction
term (\ref{eqnGuo}) was derived using a multi-scale analysis to second
order. We did not understand how these terms could be consistently derived
with a second order expansion. In a Taylor expansion the term $\nabla(\rho
FF)$ appears only as a third order term. It is likely that a second
order expansion would not pick up the $\nabla \nabla^2\rho$ term. If this
expansion can pick up the $\rho FF$ term in $\Psi$ consistently, however,
there cannot be a difference in the pre-factor of this term when it is
derived by a Taylor expansion method.

It is therefore important to compare the correction term predicted in
\cite{Guo} to our correction term $\Psi$ of (\ref{forcing2}). We
implemented the correction term (\ref{eqnGuo}). We then performed
simulations with pressure (\ref{VdW}) which does not include
interfacial terms. We would therefore expect a sharp interface. We do,
however, observe that this method leads to an extended interface,
indicating that there are additional gradient terms in the pressure
not accounted for by the Navier-Stokes equation derived by Guo
\textit{et al.}  \cite[equation (18b)]{Guo}.

We also measured the phase-diagram for the original Forcing method,
the corrections proposed by Guo and the new forcing method. The
results are shown in Figure \ref{FigGuo}. We see that the correction
introduced by Guo \textit{et al.} improves the result for the forcing
method somewhat. However, there is still no good agreement with the
theoretical phase diagram. This is, however, achieved by the correction
derived in this paper when we include additional gradient terms in the
pressure tensor (\ref{VdW_int}). This suggests that the prefactor
derived in the current work is correct.

There is a longstanding discussion about the suitability of pressure
methods for the simulation of non-ideal fluids \cite{he,Luo}. The
criticism relies on the general argument that there should be a close
correspondence between the lattice Boltzmann equation and Kinetic
Theory. At this level it is a somewhat philosophical argument. And
this philosophical argument is weakened by the occurrence of lattice
correction terms that have no analogy in Kinetic Theory. A more useful
test should be the ability of the forcing and pressure methods to
recover correct solutions in a robust and stable manner. As far as the
recovery of equilibrium solutions is concerned, we find that there is
no fundamental difference in the suitability of including non-linear
pressure terms into a lattice Boltzmann method through a forcing term
or a pressure term.

As a next step in this analysis we need to compare the performance of
the different methods under Galilean transformations. To uncover any
term that have not been previously discussed \cite{Li} requires us to
drop the assumption of small velocities in our 5th order
expansion. Those results will be published elsewhere.

To truly distinguish between the two approaches, however, the analysis
of dynamic solutions is required. There are very few tests in the
literature of non-stationary solution for liquid-gas lattice Boltzmann
simulations. But such tests will be required to compare the
differences between pressure and forcing approaches. Examples of such
simulations include advected fluids undergoing phase-separations. For
density small compared to the equilibrium densities analytical
solutions exist for $\rho(x,t)$.

In closing we want to point out that the corrections for the forcing
method for phase-separated systems do also apply for external
forces. The same additional pressure terms that we identified in
(\ref{Peff}) also occur when a truly external force acts on the
system. This is particularly important to keep in mind when simulating
liquid-gas systems in the presence of gravity.

\end{document}